\documentclass[twocolumn,superscriptaddress,floatfix,longbibliography,aps,pra,nopreprintnumbers]{revtex4-2}

\usepackage[utf8]{inputenc}
\usepackage[T1]{polski}
\usepackage[english]{babel}
\usepackage{graphicx}
\usepackage{bm}
\usepackage{bbm}
\usepackage{xcolor}
\usepackage{amsmath}
\usepackage{amssymb}
\usepackage[normalem]{ulem} %strikethrough \sout
\usepackage{enumerate}
\usepackage[urlcolor=blue,colorlinks=true,citecolor=blue,linkcolor=blue,pdfstartview={FitH},bookmarks=false]{hyperref}
\graphicspath{{Figures/}{./Figures/}{.}}

\allowdisplaybreaks

\begin{document}

\title{Information Trapping by Topologically Protected Edge States: Scrambling and the Butterfly Velocity}
\author{Martyna Sedlmayr}
\affiliation{Lublin University of Technology, Faculty of Mechanical Engineering,
	Nadbystrzycka 31, 20-618 Lublin, Poland}
\affiliation{Polish Air Force University, Faculty of Aviation, Dywizjonu 303 St 35, 08-521 D\k{e}blin, Poland}
\author{Hadi Cheraghi}
\affiliation{Institute of Physics, Maria Curie-Sk\l{}odowska University, 20-031 Lublin, Poland}
\author{Nicholas Sedlmayr}
\email[e-mail: ]{sedlmayr@umcs.pl}
\affiliation{Institute of Physics, Maria Curie-Sk\l{}odowska University, 20-031 Lublin, Poland}

\date{\today}

\begin{abstract}
Topological insulators and superconductors have recently attracted considerable attention, and many different theoretical tools have been used to gain insight into their properties. Here we investigate how perturbations can spread through exemplary one-dimensional topological insulators and superconductors using out-of-time ordered correlators. Out-of-time ordered correlators are often used to consider how information becomes scrambled during quantum dynamics. The wavefront of the out-of-time ordered correlator can be ballistic regardless of the underlying system dynamics, and here we confirm that for {\it topological} free fermion systems the wavefront spreads linearly at a characteristic butterfly velocity. We pay special attention to the topologically protected edge states, finding that ``information'' can become trapped in the edge states and essentially decoupled from the bulk, surviving for relatively long times. We then generalise this to consider several different models with multiple possible edge states coexisting on a single edge.
\end{abstract}

\maketitle

%%%%%%%%%%%%%%%%%%%%%%%%%%%%%%%%%%%%%%%%%%%%%
\section{Introduction}
%%%%%%%%%%%%%%%%%%%%%%%%%%%%%%%%%%%%%%%%%%%%%

Advances in experimentally controlling matter at the quantum level have enabled the measuring of the dynamics of quantum information in real-time \cite{Cheneau2012,Landsman2019}, resulting in a rise in questions about how information can spread in many-body systems. In classical physics, the hallmark of chaotic systems is characterized by the exponential divergence in phase space of states initially close together, bound by the Lyapunov exponent. A quantum Lyapunov exponent $\lambda_L$ is found to bound the growth of information scrambling in a quantum system characterized by an out-of-time-ordered commutator (OTOC)~\cite{Maldacena2016}. This scrambling spreads through the system at a so-called butterfly velocity $v_b$~\cite{Maldacena2016}, which can be understood as an effective Lieb-Robinson bound for the state of the quantum system~\cite{Lieb1972,Roberts2016}. Quantum scrambling is often considered a quantum analog of chaotic dynamics in classical systems which, although the dynamics are ultimately unitary and therefore reversible, encodes the loss of information across the degrees of freedom of the system~\cite{Chowdhury2017}, though these concepts should be taken as distinct~\cite{Xu2020b}. Scrambling can occur at exponentially slow~\cite{McGinley2019b} or fast~\cite{Belyansky2020} rates. In addition to quantum chaos, another fundamental question in quantum dynamics is under what circumstances a system thermalizes~\cite{Sirker2014a,Deutsch2018}, and scrambling has also been applied to help understand thermalization and entanglement growth~\cite{VonKeyserlingk2018,Alba2019,Modak2020}.

An OTOC can be understood as a two-time correlation function in which operators are not chronologically ordered in time. First introduced in the context of superconductivity~\cite{Larkin1969} they are now widely studied in quantum chaos and quantum dynamics more generally. Due to the feasibility of measuring the OTOC experimentally~\cite{Swingle2016,Garttner2017,Li2017,Lewis-Swan2019,Nie2020,Blok2021,Sundar2022,Weinstein2022}, it has attracted a lot of interest in physics across many different fields, from statistical physics and thermodynamics~\cite{Chenu2018,Campisi2017}, through conformal field theories and black holes~\cite{Patel2017,Cotler2017}, to Luttinger liquids and quantum impurity systems~\cite{Dora2017,Dora2017a}. There are a multitude of examples of the uses of OTOCs. These include being used as a tool to detect many-body localized systems as potentially less chaotic or as slow scramblers~\cite{Swingle2017,Slagle2017} and to analyze dynamically ergodic-nonergodic transitions~\cite{Buijsman2017}. Additionally, they are applied to excited-state quantum phase transitions~\cite{Wang2019d}, equilibrium phase transitions~\cite{Dag2019}, as well as dynamical quantum phase transitions~\cite{Heyl2018,Heyl2018a}.  OTOCs have been applied to a wide variety of models such as the Sachdev-Ye-Kitaev model~\cite{Maldacena2016a}, the $O(N)$ model~\cite{Chowdhury2017}, disordered systems~\cite{Swingle2017}, models with critical Fermi surfaces and hard-core boson models~\cite{Patel2017, Lin2018a}, random field XX spin chains~\cite{Riddell2019}, and Floquet quantum systems~\cite{Zamani2022}.
 
While most recent studies are based on bulk quantum systems (typically with periodic boundary conditions), comparatively little attention has been focused on open quantum systems and boundary effects. For topological matter~\cite{Hasan2010} the boundaries are of great interest due to the existence of the topologically protected edge states, guaranteed to exist due to the bulk boundary correspondence~\cite{Hasan2010,Teo2010}. Here we focus on one-dimensional symmetry-protected topological systems~\cite{Chiu2016}. Depending on the symmetries these systems can have either $\mathbb{Z}_2$ or $\mathbb{Z}$ topological invariants~\cite{Schnyder2009,Ryu2010} and hence either a single edge state or multiple edge states existing at a boundary respectively. We focus on two exemplary systems, the Su-Schrieffer-Heeger (SSH) chain~\cite{Su1980,Sirker2014} and a generalization of the Kitaev chain with longer range coupling terms~\cite{Kitaev2001,Sedlmayr2018}. Both of these are in the symmetry class $BDI$ and can therefore potentially possess multiple protected boundary states. However we focus on particular models where the SSH chain has a topological index of 0 or 1, and the Kitaev chain of 0,1,2, or 3. Such topological insulators and superconductors have been shown to undergo dynamical quantum phase transitions~\cite{Heyl2013,Heyl2018a,Sedlmayr2019a} following appropriate quenches~\cite{Vajna2015,Sedlmayr2018,Uhrich2020} and to posses a dynamical bulk-boundary  correspondence~\cite{Sedlmayr2018,Sedlmayr2019a,Maslowski2020,Maslowski2023}. As the robust zero-energy edge modes are strongly robust to disorder~\cite{Mittal2014,Meier2018} and defects~\cite{Bandres2018}, they are applied to realize robust transport~\cite{St-Jean2017,Zhao2018,Parto2018}, topologically protected quantum coherence~\cite{Bahri2015,Nie2020a}, and quantum state transfer~\cite{Yao2013}. It is therefore of fundamental interest to seek the effects of edge modes on information scrambling.

OTOCs have been applied to study the topological phase transition~
\cite{Dag2019,Bin2023} and to scrambling in spin chains with topological order \cite{Orito2022}, as well as to topological Floquet systems~\cite{Sur2022a}. Here we consider scrambling in topological insulators and superconductors, focusing on the role of the boundary and any topological-protected boundary modes. We find that if the time-evolving Hamiltonian is topologically non-trivial, and hence has boundary modes associated with it, then information becomes trapped in the edge mode. By applying quench protocols we find that the initial state of the system is not key to how information spreads, nor to the behavior of the boundary modes. The short time scrambling for the SSH model can be fitted with a Lyapunov exponent and butterfly velocity. However for a Kitaev chain which includes next-next-nearest neighbor hopping and p-wave pairing terms, the fits become significantly worse. Nonetheless, a butterfly velocity can be clearly seen in the results. Beyond the boundary effects, we see no further effect of the topological order on the scrambling. For perturbations at the edge modes, the scrambling  propagates via the bulk as expected. 

This paper is organized as follows. In Sec.~\ref{sec:scrambling} we introduce the OTOCs and the methods we use to calculate the results of the article, then in Sec.~\ref{sec:models} we define the exemplary models we study and the perturbations used. Secs.~\ref{sec:ssh} and \ref{sec:lrk} summarise the results of the calculations and finally we conclude in Sec.~\ref{sec:concl}.

%%%%%%%%%%%%%%%%%%%%%%%%%%%%%%%%%%%%%%%%%%%%%
\section{OTOCs and non-translationally invariant systems}\label{sec:scrambling}
%%%%%%%%%%%%%%%%%%%%%%%%%%%%%%%%%%%%%%%%%%%%%

We will consider the spread of perturbations to the system by using the out-of-time ordered commutator and correlator. Let us consider two unitary operators $\hat V_j$ and $\hat W_j$ describing local perturbations to a lattice model at site $j$. We will consider perturbations that both break and preserve the symmetries important for the topological order. The time evolution of $\hat W_j$ under some Hamiltonian $\hat H$ is given by $\hat W_j(t)=e^{i \hat Ht}\hat W_je^{-i \hat Ht}$. The out-of-time ordered commutator (OTOC) can be defined as~\cite{Swingle2016,Roberts2016}
\begin{equation}\label{c1}
	C_j(t)=\left\langle\left[\hat W_{j_0}(t),\hat V_j\right]^\dagger\left[\hat W_{j_0}(t),\hat V_j\right]\right\rangle\,.
\end{equation}
This definition, which we will focus on, has the advantage of being Hermitian and therefore having only real (and in this case positive) eigenvalues. The site of the perturbation $\hat W_{j_0}$ is taken as a parameter of the model and we look at the dependence of the commutator on the location of $\hat V_j$. An alternative definition which is also often used is~\cite{Larkin1969,Dora2017a,Heyl2018} $C_{\rm L}(t)= -\left\langle\left[\hat W_{j_0}(t),\hat V_j\right]^2\right\rangle$.
We can rewrite $C_j(t)$ as $C_j(t)=2(1-\Re[F_j(t)])$ with $F_j(t)$ the out-of-time ordered correlator
\begin{equation}
	F_j(t)=\left\langle \hat W_{j_0}^\dagger(t)\hat V_j^\dagger \hat W_{j_0}(t)\hat V_j\right\rangle\,.
\end{equation}
We will now focus on how we calculate $F_j(t)$ for our models in the absence of translational invariance. The average $\langle\ldots\rangle$ can be over any state of interest $\left|\psi\right\rangle$. Here we mostly focus on the case where $\left|\psi\right\rangle$ is the ground state of $\hat H$. Taking a different ground state does not affect the results, see appendix \ref{app:quenches}. The physical intuition behind this correlator is that it is measuring the impact of one observable at earlier times on another observable at later times.

Let us start the system in the state $|\psi_0\rangle$. Therefore all averages are here taken to be $\langle\ldots\rangle=\langle\psi_0|\ldots|\psi_0\rangle$. It is already known that the Loschmidt echo \cite{Levitov1996,Klich2003,Rossini2007,Sedlmayr2018} can be rewritten as the determinant of a matrix defined in terms of the correlation matrix,
\begin{equation}
{\bm M}=\langle\psi_0|\hat\Phi^\dagger\hat\Phi|\psi_0\rangle\,,
\end{equation} 
where $\Phi$ is the single particle annihilation operator written in an appropriate basis: $\hat\Phi=(\hat c_1,\hat c_2,\ldots)$. In exactly the same way one can find
\begin{equation}\label{eq:f}
	F_j(t)=\det\left[1+ {\bm M}\left( {\bm W}_{j_0}^\dagger(t){\bm V}_j^\dagger {\bm W}_{j_0}(t){\bm V}_j-1\right)\right]\,.
\end{equation}
In this case, ${\bm W}_{j_0}(t)$ and ${\bm V}_j$ are understood to be written in the same basis as ${\bm M}$. Eq.~\ref{eq:f} is one of the important results of our article, allowing an efficient calculation of the OTOCs for free fermion systems with open boundary conditions.

At short times in an integrable quantum system, it is expected that the OTOC follows an exponential increase given by~\cite{Lin2018,Xu2020a,Xu2022}
\begin{equation}\label{cfit}
    C_j(t)\sim e^{\lambda_L\frac{\left(t-\frac{a|j-j_0|}{v_b}\right)^{3/2}}{t^{1/2}}}\,.
\end{equation}
$\lambda_L$ is the Lyapunov exponent and $v_b$ the butterfly velocity. We use this to gain an estimate for the butterfly velocity which generally gives a good fit for our results.

%%%%%%%%%%%%%%%%%%%%%%%%%%%%%%%%%%%%%%%%%%%%%
\section{Exemplary one-dimensional topological models}\label{sec:models}
%%%%%%%%%%%%%%%%%%%%%%%%%%%%%%%%%%%%%%%%%%%%%

Let us now introduce the two exemplary one-dimensional topological models which we will consider in the rest of the article. Both of these are two band models which can be written generically in momentum space as
\begin{equation}
	\hat H=\sum_{ k}\hat \Psi^\dagger_{ k}\mathcal{H}( k)\hat \Psi_{ k} \textrm{ where } \mathcal{H}(k)=\mathbf{d}_k \cdot {\bm\tau}\,,
\end{equation}
where ${\bm\tau}$ are Pauli matrices in some physical subspace, $\hat \Psi_{ k}$ is a fermionic operator in the same subspace, and $\mathbf{d}_k=(d^x_k,d^y_k,d^z_k)$ parameterizes the Hamiltonian density $\mathcal{H}( k)$. These Hamiltonians have pairs of eigenenergies $\pm \epsilon_k$, due to the particle-hole symmetry of the Hamiltonians. Additional symmetries apply further constraints on $\mathbf{d}_k$. For example, the models we consider have a unitary chiral symmetry $\mathcal{S}$, given by $\left[\mathcal{S},\mathcal{H}(k)\right]_+=0$, which places them in class BDI and allows for the calculation of a $\mathbb{Z}$ topological invariant:
\begin{equation}
    \nu=\frac{1}{4\pi i}\int dk\, \textrm{tr}\, \mathcal{S}
    \left[\partial_k\mathcal{H}(k)\right]\left[\mathcal{H}(k)\right]^{-1}\,.
\end{equation}

The extensively studied SSH chain~\cite{Su1980,Sirker2014a} is a simple lattice with
alternating hopping strengths $J(1\pm\delta)$. The
Hamiltonian is
\begin{align}
\label{RM_model}
	\hat H_{\rm SSH}=&-J(1+\delta)\sum_{j=1}^{N/2}\hat \Psi^\dagger_j{\bm \sigma}^x
    \hat \Psi_j\\\nonumber&-J(1-\delta)\sum_{j=1}^{N/2-1}\hat \Psi^\dagger_j
\begin{pmatrix}0&0\\1&0\end{pmatrix}\hat \Psi_{j+1}+\textrm{H.c.}\,,
\end{align}
where $\hat \Psi^\dagger_{j}=(\hat a^\dagger_{j},\hat b^\dagger_{j})$. $\hat a_{
j}^{\dagger}$ and $\hat b_{
j}^{\dagger}$ create spinless fermionic particles at the two sublattice sites in unit cell $j$. One then finds
\begin{equation}
	\mathbf{d}_k=\begin{pmatrix}
	-2J \cos k, 2J\delta\sin k,0
    \end{pmatrix}\,.
\end{equation}
Here the physical subspace is the basis of the unit cell. The SSH chain can be written more straightforwardly directly in real space as
\begin{equation}
	\hat H_{\rm SSH}=-J\sum_{j=1}^{N-1}\left[(1+\delta e^{i\pi j})\hat c^\dagger_j\hat c_{j+1}+\textrm{H.c.}\right]\,,
\end{equation}
where $\hat a_{j}=\hat c_{2j-1}$ and $\hat b_{j}=\hat c_{2j}$. This is the version we will use in the following.

\begin{figure}
	\includegraphics[width=0.9\columnwidth]{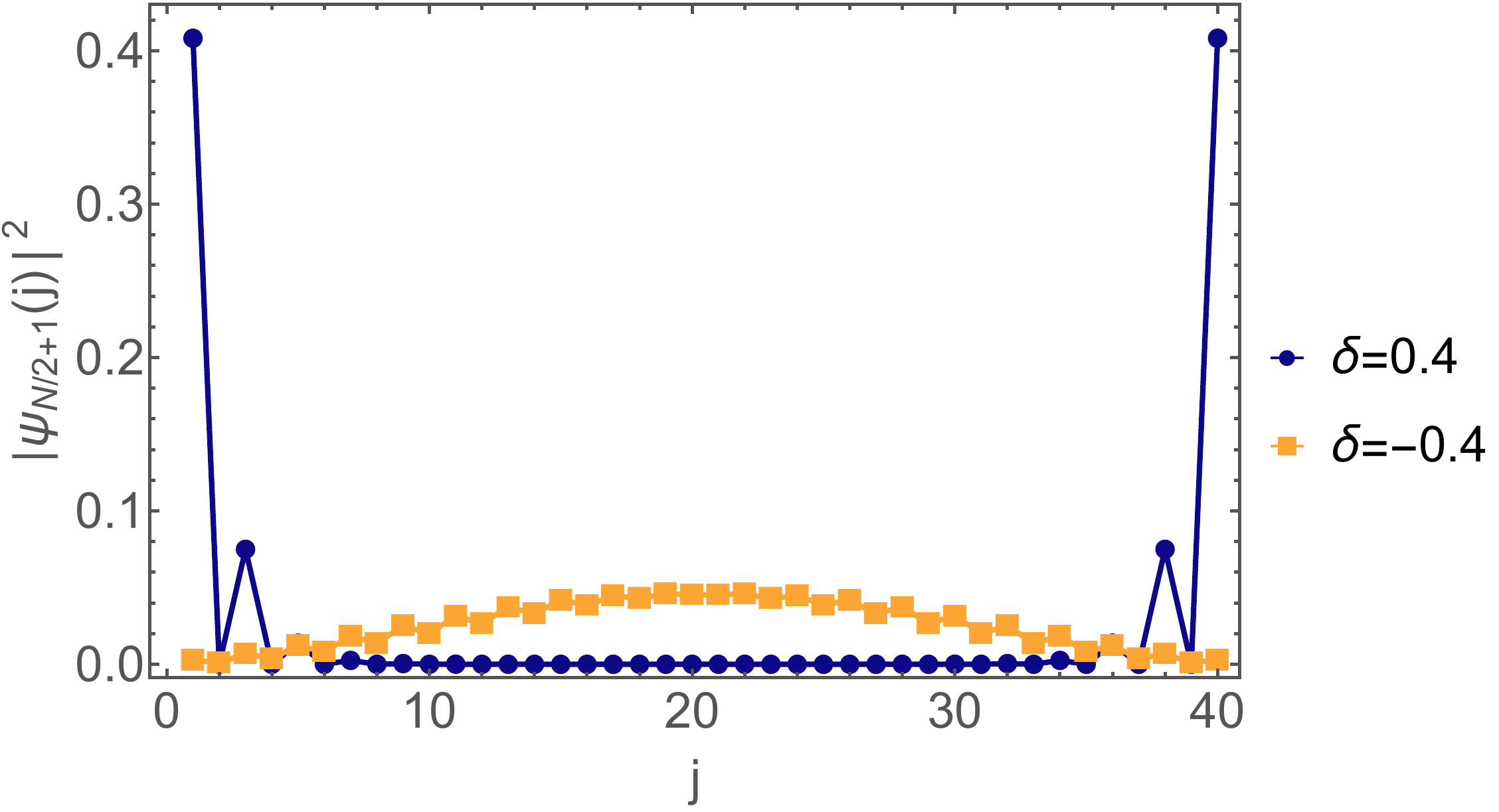}
	\caption{Density of the lowest positive energy eigenstates of the SSH model with $\delta=\pm0.4$, $N=40$. At these parameters, the edge states in the topologically non-trivial phase can be seen to be well localized across the first and third sites from the edges, with effectively zero overlap between the edges. For the topologically trivial phase, the lowest positive energy eigenstate is of course a bulk state.}
	\label{fig_ssh_density}
\end{figure}

For the SSH model, a natural perturbation is an onsite density term and so we take the unitary operators $\hat W_{j_0}=\exp\left(i W_0\hat c^\dagger_{j_0}\hat c_{j_0}\right)=\exp\left(i W_0\hat n_{j_0}\right)$ and $\hat V_j=\exp\left(i V_0\hat c^\dagger_j\hat c_j\right)=\exp\left(i V_0\hat n_j\right)$. Throughout the paper, we will consider the values $V_0=W_0=5$. No results depend on these values and we find good graphical demonstrations for these values. The system size is taken to be $N=40$, which again is picked to be sufficient to show the effects we focus on. In Fig.~\ref{fig_ssh_density} we show examples of the lowest positive energy wavefunction for $\delta=\pm0.4$. The dimerization is taken large enough to avoid any overlap between zero modes on the opposite sides of the chain. We note that increasing the chain length decreases the zero mode energy exponentially, which can result in multiple precision being necessary for the numerical solutions~\cite{Sirker2014a}.

\begin{figure}
	\includegraphics[width=0.9\columnwidth]{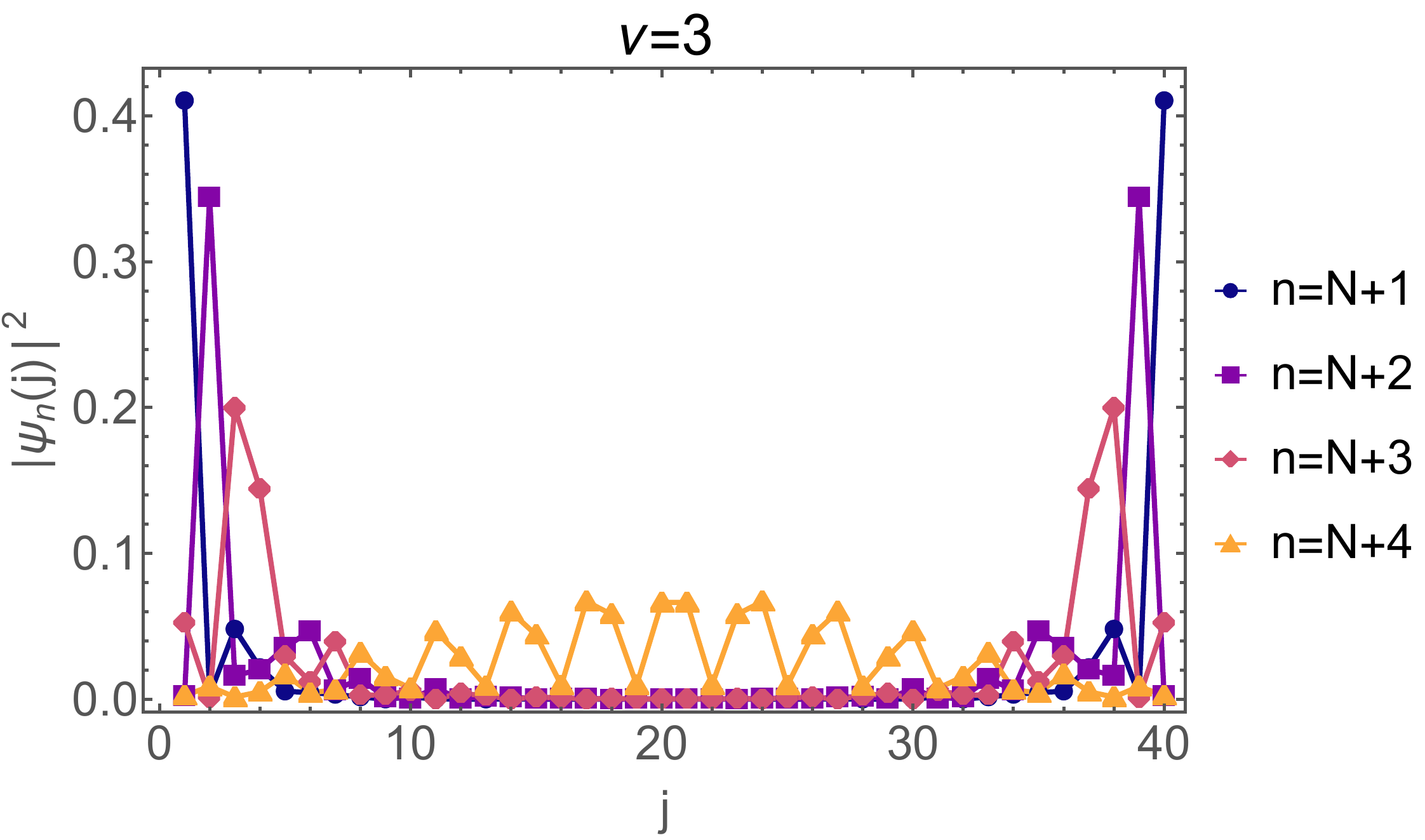}
	\caption{Density of the lowest four positive energy eigenstates of the Kitaev chain in the $\nu=3$ topological phase, see Table \ref{tab:quenches}, which has three zero modes. The chain length is $N=40$. For this case, and the other cases in Table \ref{tab:quenches}, all edge states are well localized at the edges, and the lowest positive energy bulk state is also shown.}
	\label{fig_kit_density}
\end{figure}

A related model is a long-range Kitaev chain~\cite{Kitaev2001,DeGottardi2013,Sedlmayr2018}, where we include hopping of up to a distance of $R$ lattice spacings. This Hamiltonian can be written as 
\begin{align}
	\hat H_{\rm K}=&\sum_{j}\sum_{\ell=1}^R\Psi^\dagger_{j}\left(
	\Delta_{\ell}i{\bm\tau}^y-J_{\ell}{\bm\tau}^z\right)\Psi_{j+\ell}+\textrm{H.c.}\nonumber\\&-\mu\sum_{j}\Psi^\dagger_{j}{\bm\tau}^z\Psi_{j}\,,
\end{align}
where the operators in particle-hole space are
given by $\hat \Psi^\dagger_{j}=(\hat a^\dagger_{j},\hat a_{j})$. $\hat a_{
j}^{(\dagger)}$ annihilates (creates) a spinless fermionic particle at
a site $j$. For the hopping we have $\vec J=(J_1,J_2,\ldots J_R)$ and for the pairing
$\vec\Delta=(\Delta_1,\Delta_2,\ldots \Delta_R)$, $\mu$ is the chemical potential. This results in
\begin{equation}
	\mathbf{d}_k=\sum_{\ell=1}^R \begin{pmatrix}
	-2J_\ell\cos[\ell k]-\mu/R,2\Delta_\ell\sin[\ell k],0
\end{pmatrix}\,.
\end{equation}
Here we will focus on $R=3$, which allows us to tune between different topological phases beyond $\nu=0$ and $\nu=1$. $J_\ell$ and $\Delta_m$ are taken to be arbitrarily tunable, rather than having realistic values, in order to reach convenient phases. Throughout this paper we take energies to be measured in terms of a hopping strength $J$ and we set $\hbar=1$. As for the SSH chain, we also take $N=40$. In Table \ref{tab:quenches} we list the parameters used for each topological phase we consider. In Fig.~\ref{fig_kit_density} we give examples of the lowest positive energy eigenvalues of the wavefunctions in the phase $\nu=3$, just as for the case of the SSH chain it is important that these states do not overlap.

For the Kitaev chain, there are a variety of natural local perturbations which can be considered, as well as many alternative possibilities. We mainly focus on $\hat W_{j_0}=\exp\left(i W_0\hat \Psi^\dagger_{j_0}{\bm\tau}^\alpha\hat \Psi_{j_0}\right)$ and $\hat V_j=\exp\left(i V_0\hat \Psi^\dagger_j{\bm\tau}^\alpha\hat \Psi_j\right)$ As for the SSH chain we will consider the values $V_0=W_0=5$. For the perturbation, we can take $\hat V_j$ and $W_{j_0}$ to have the same $\alpha$ or different ones. We can also take perturbations that either do or do not respect the symmetries of the Kitaev chain. In general, we find no dependence of the results on these variations and just show a few select examples in the following. We also tested more unusual perturbations which also show similar behavior.

\begin{table}[]
    \begin{center}
    \begin{tabular}{|c|c|c|c|}
        \hline $\nu$ & $\vec{J}/J$ & $\vec{\Delta}/J$ & $\mu/J$\\\hline
        0 & $(1,0.5,-0.25)$ & $(1.2,0.6,0.3)$ & 3 \\\hline
        1 & $(1,-2,2)$ & $(1.3,-0.6,0.6)$ & 2 \\\hline
        2 & $(1,-2,0)$ & $(0.45,-0.9,0)$ & 0.1 \\\hline
        3 & $(1,-2,2)$ & $(0.45,-0.9,1.35)$ & 0.1 \\\hline
    \end{tabular}\caption{The parameter values for the exemplary points chosen in the topological phase diagram of the Kitaev chain. All results in this article refer to these parameter values.}
    \label{tab:quenches}
    \end{center}
\end{table}

%%%%%%%%%%%%%%%%%%%%%%%%%%%%%%%%%%%%%%%%%%%%%
\section{Results for the SSH model}\label{sec:ssh}
%%%%%%%%%%%%%%%%%%%%%%%%%%%%%%%%%%%%%%%%%%%%%

We first focus on the SSH model, which amply demonstrates the basic effects. For all results here the initial state $|\psi_0\rangle$ is taken to be the half-filled groundstate of the time-evolving Hamiltonian $\hat H$. We look at an example where $\hat H$ is in the topologically non-trivial phase ($\delta=0.4$) where it has a pair of edge states and in the topologically trivial phase ($\delta=-0.4$). Similar results are seen for other values. We also investigated quenches, in which $|\psi_0\rangle$ and $\hat H$ belong to different topological phases. However, we find that only $\hat H$ controls the interesting behavior. The initial state plays a secondary role, see App.~\ref{app:quenches} for more details.

\begin{figure}
	\includegraphics[width=0.49\columnwidth]{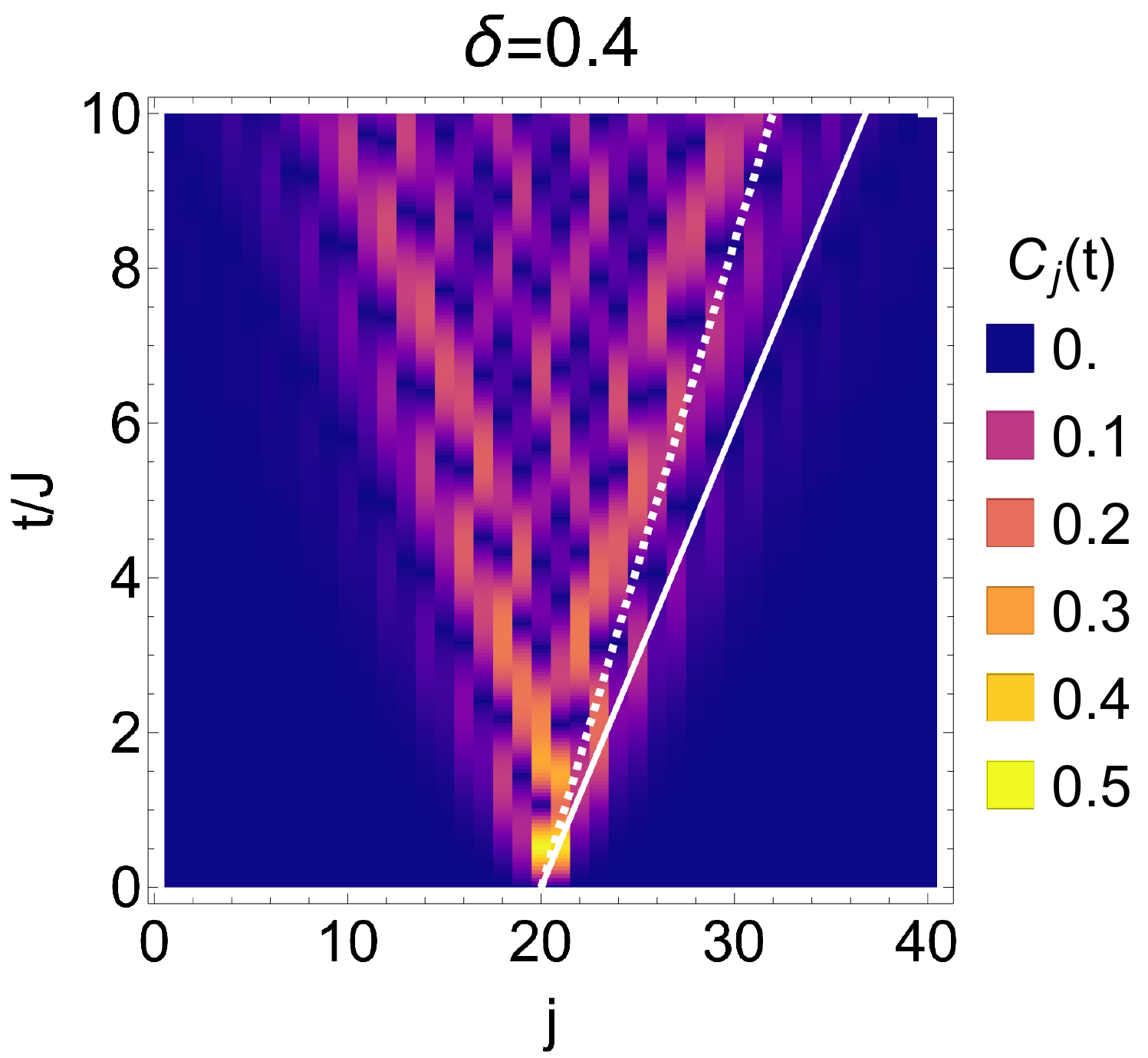}
	\includegraphics[width=0.49\columnwidth]{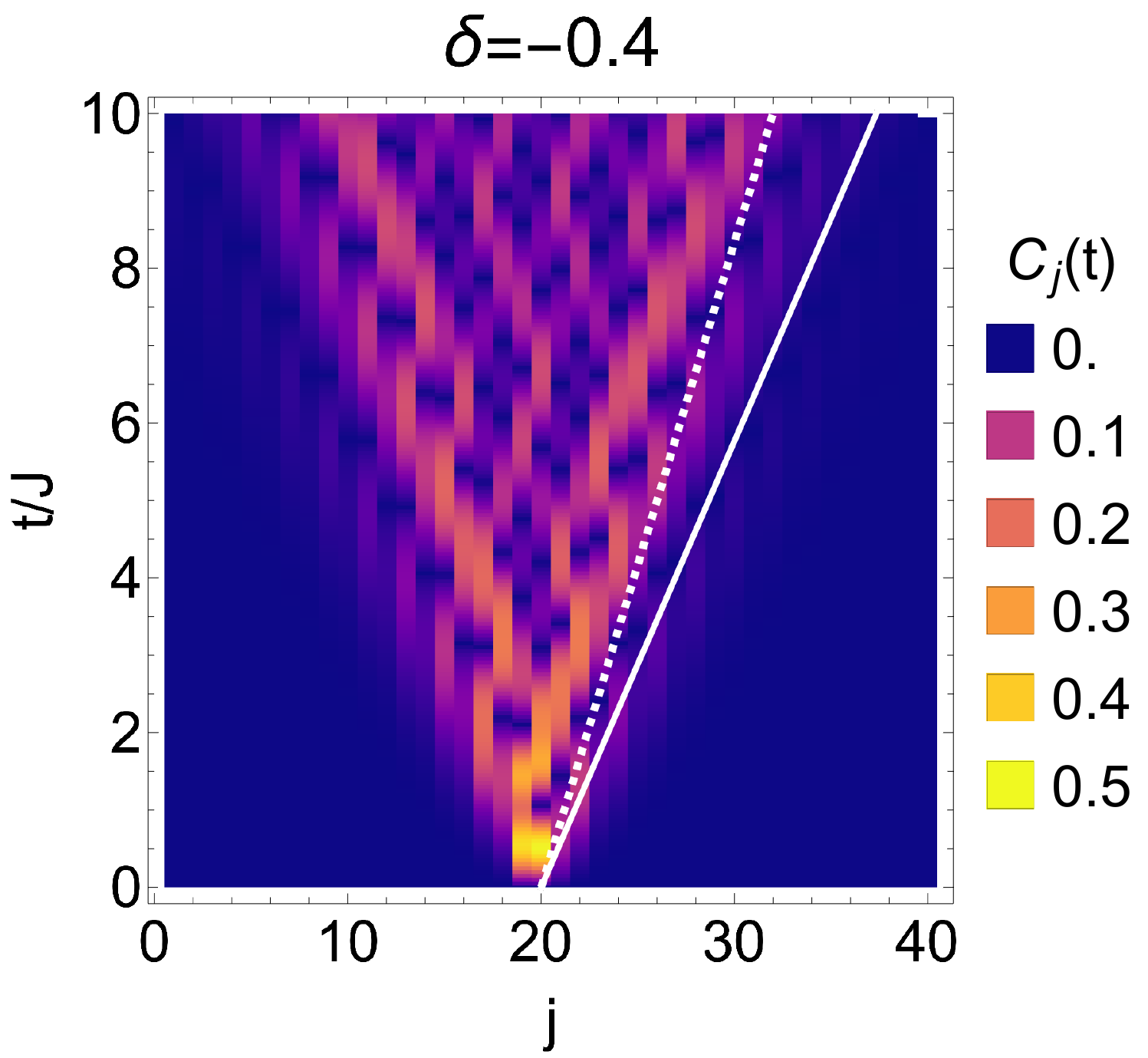}\\
	\includegraphics[width=0.49\columnwidth]{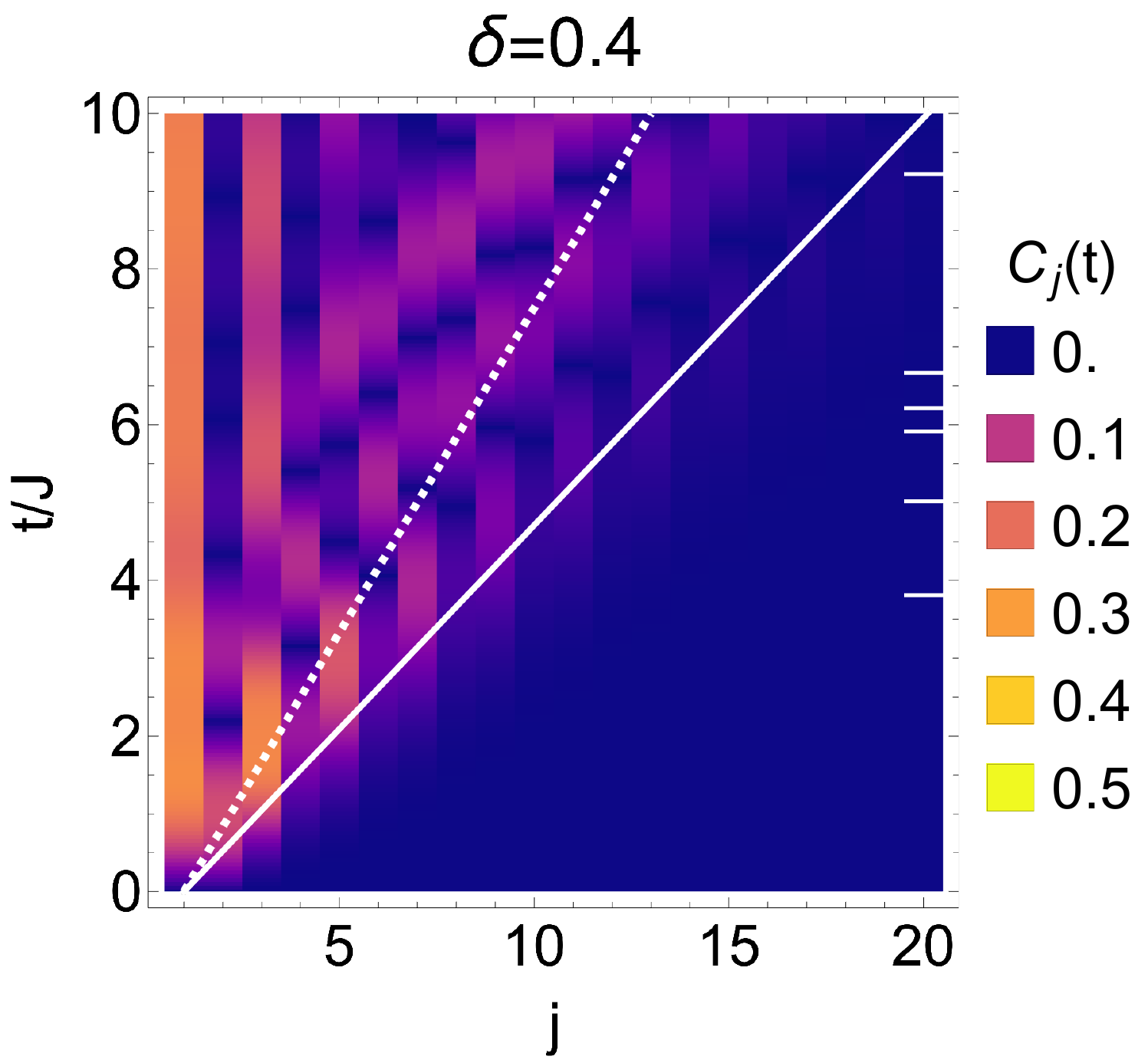}
	\includegraphics[width=0.49\columnwidth]{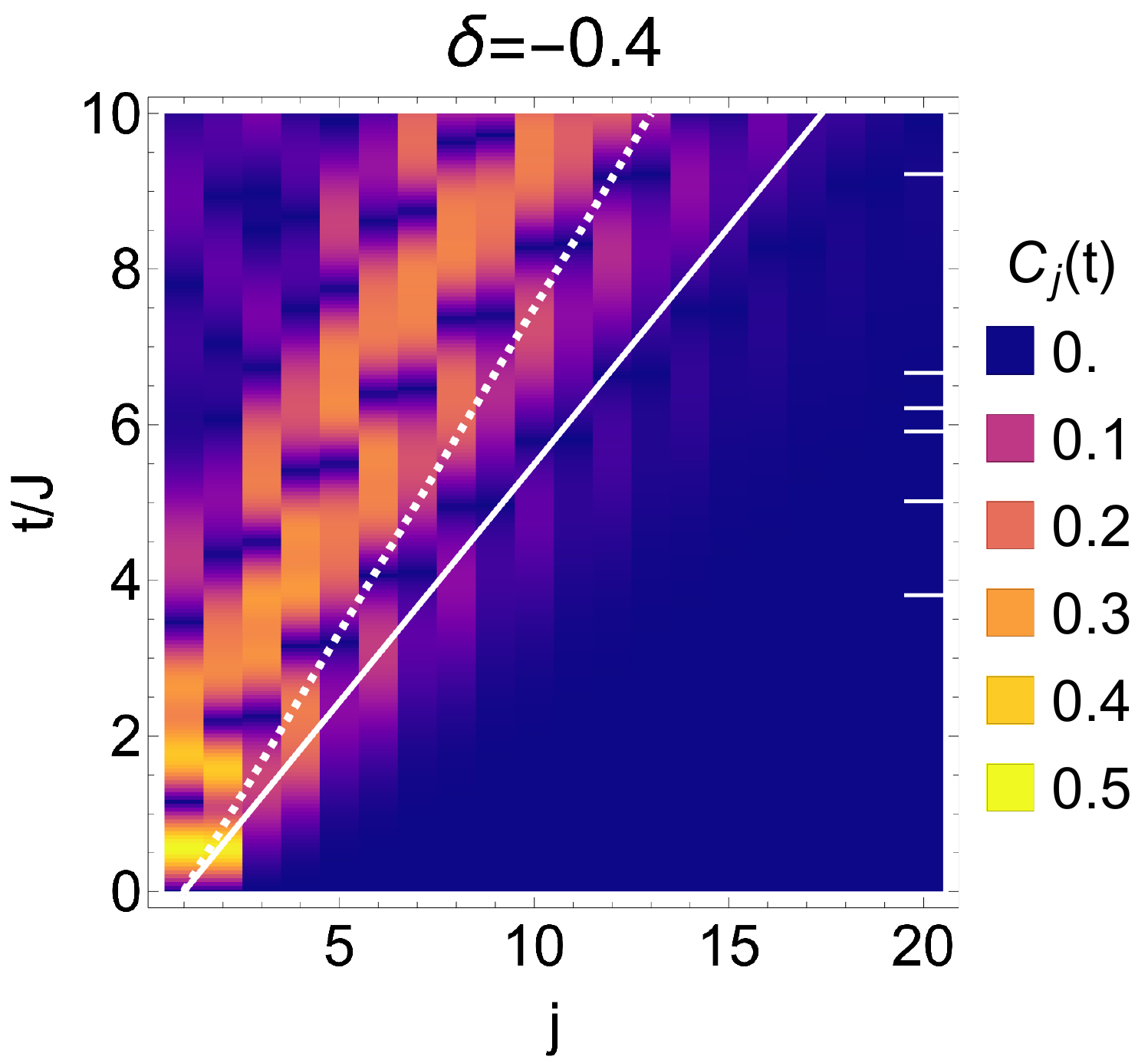}
	\caption{The OTOC $C_j(t)$ for the SSH model. The perturbation occurs at $j_0=20$ in the upper panels and $j_0=1$ in the lower panels and the system size is $N=40$. The white lines are an aid to the eye for how fast the correlations spread, they show the maximum group velocity of the bulk bands $v_g=2J(1-|\delta|)$ (dashed lines) and the butterfly velocity $v_b$ (solid lines) found from fits to Eq.~\eqref{cfit}, see App.~\ref{app:fits}.}
	\label{figssh1}
\end{figure}

In Fig.~\ref{figssh1} $C_j(t)$ is plotted for the two different topological phases and for $j_0=1$ and $j_0=20$. I.e.~for a perturbation at the edge and in the center of the chain. In the bulk little effect can be seen from which phase is being considered. The perturbation spreads as expected with a constant velocity, the butterfly velocity. Plotted as a comparison are two velocities. One is the butterfly velocity extracted from a fit to Eq.~\eqref{cfit}, see App.~\ref{app:fits} for details. The second is the maximum group velocity of the equilibrium bands. For the SSH model, this is easily calculated and results in $v_g=2J(1-|\delta|)$. These velocities capture some of the spread of correlations in the system, however, they clearly do not give an absolute bound. There is an odd-even effect in the spread of the correlations, which we will look at in more detail below. As the two bulk topological phases are equivalent for $\delta\to-\delta$ after reflection about a site, one can see that the bulk spreads are also symmetric. A very clear difference between the phases can be seen only at the boundary. When $\hat H$ belongs to the topologically non-trivial phase, and therefore has topologically protected edge modes, information becomes trapped in these modes at the edge of the system.

\begin{figure}
	\includegraphics[width=0.9\columnwidth]{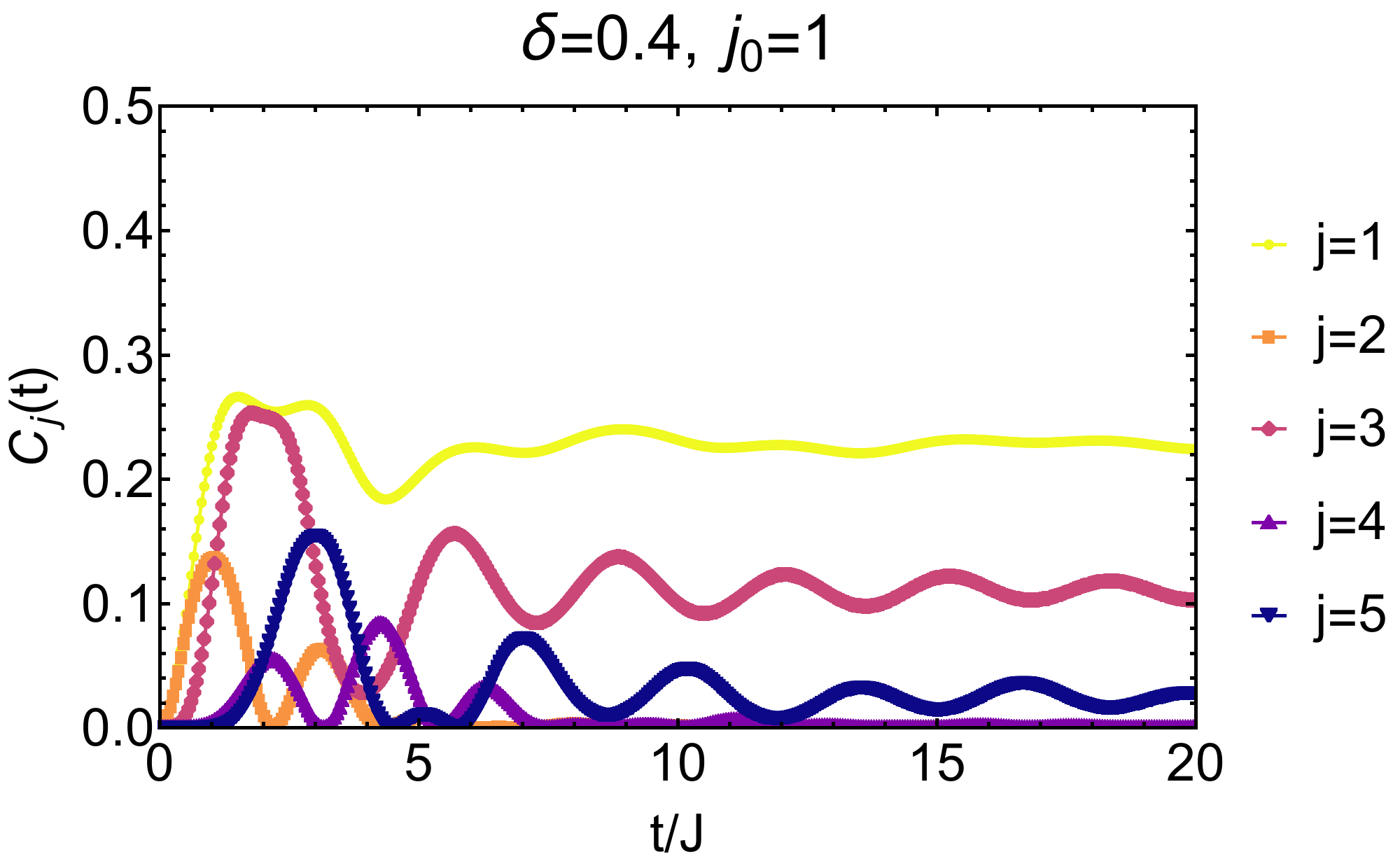}\\
 	\includegraphics[width=0.9\columnwidth]{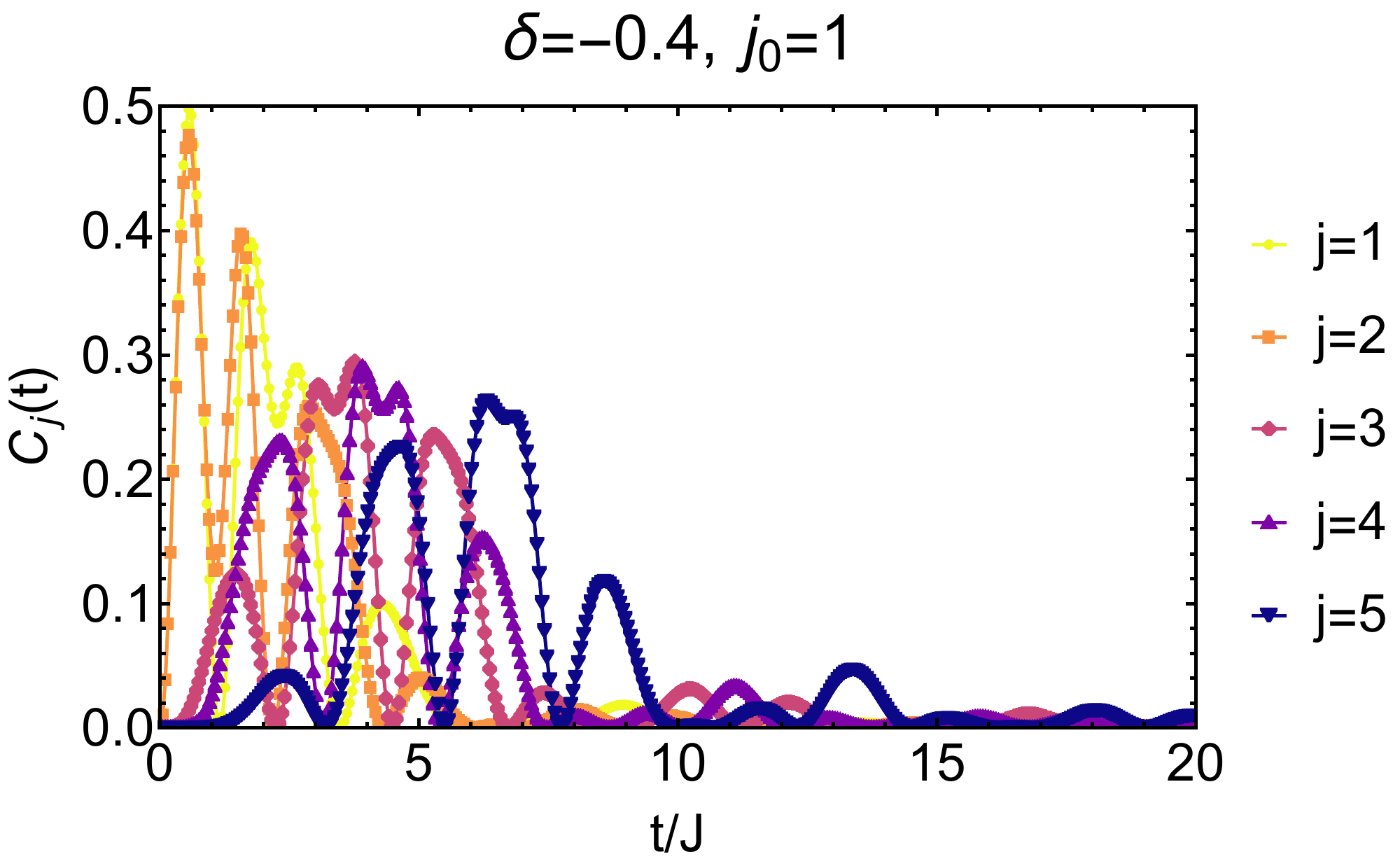}
	\caption{The OTOC $C_j(t)$ for the SSH model near the boundary. The perturbation occurs at $j_0=1$ and the system size is $N=40$. Results for the topologically non-trivial phase $\delta=0.4$ and the trivial phase $\delta=-0.4$ are compared.}
	\label{figsshpos1}
\end{figure}

If we consider just the dynamics of $C_j(t)$ on several sites then this effect becomes perhaps even clearer. Fig.~\ref{figsshpos1} shows these two cases, exactly as for the lower panels of Fig.~\ref{figssh1}. When topologically protected edge modes are present, $\delta=0.4$, information is trapped in this mode. We note that this trapping follows the same even-odd effect as the density for the edge mode, see Fig.~\ref{fig_ssh_density}. Information is only trapped where the edge mode has non-zero density. By comparison when there are no edge modes one sees only the short-time transient correlations. These look rather messy due to the scattering from the open boundary. A similar effect can be seen whereby quantum coherence is long lived at the edges of one dimensional systems which have topological boundary modes or spin chains with localized strong zero modes~\cite{Haque2010,Alba2013,Kemp2017}.

\begin{figure}
	\includegraphics[width=0.49\columnwidth]{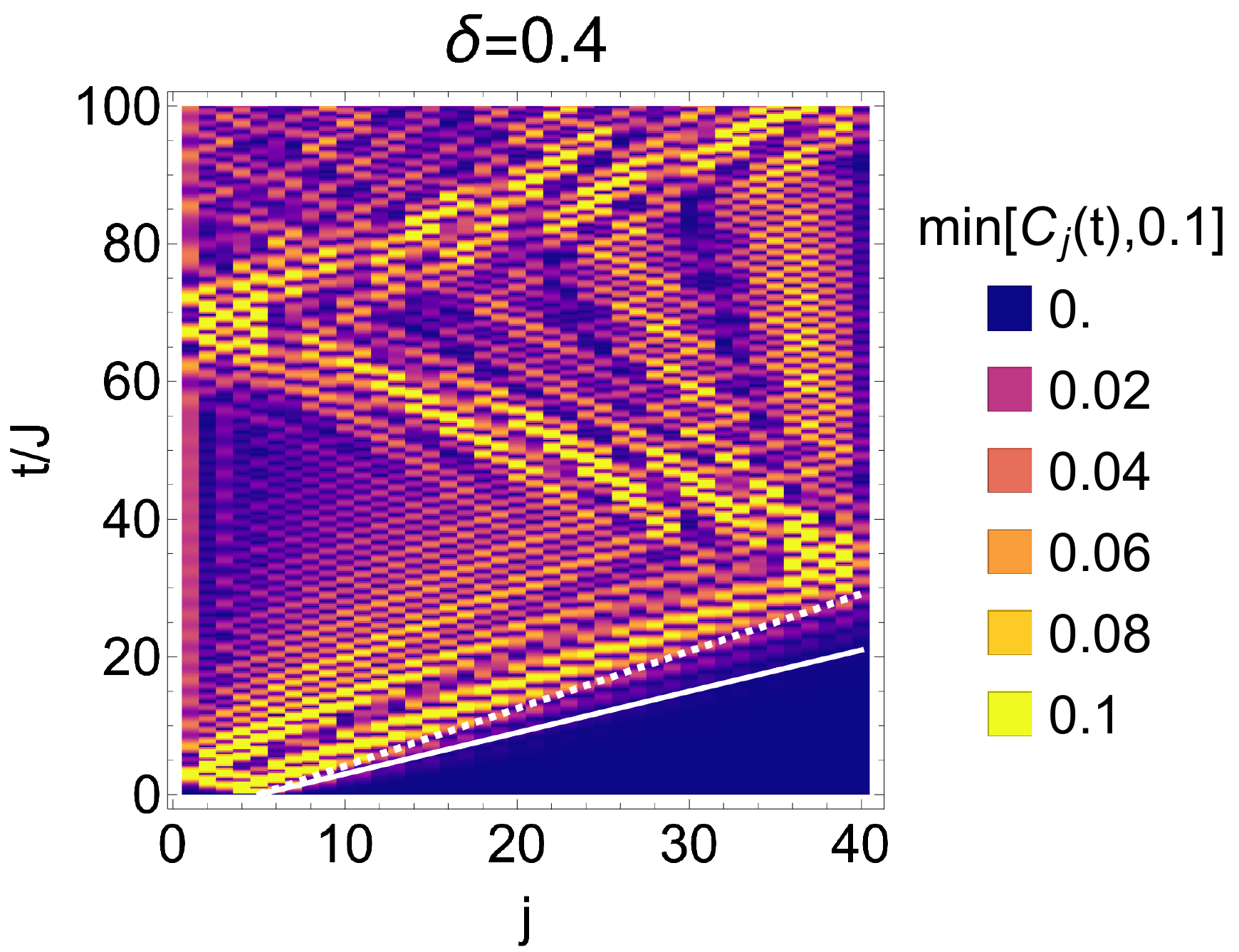}
	\includegraphics[width=0.49\columnwidth]{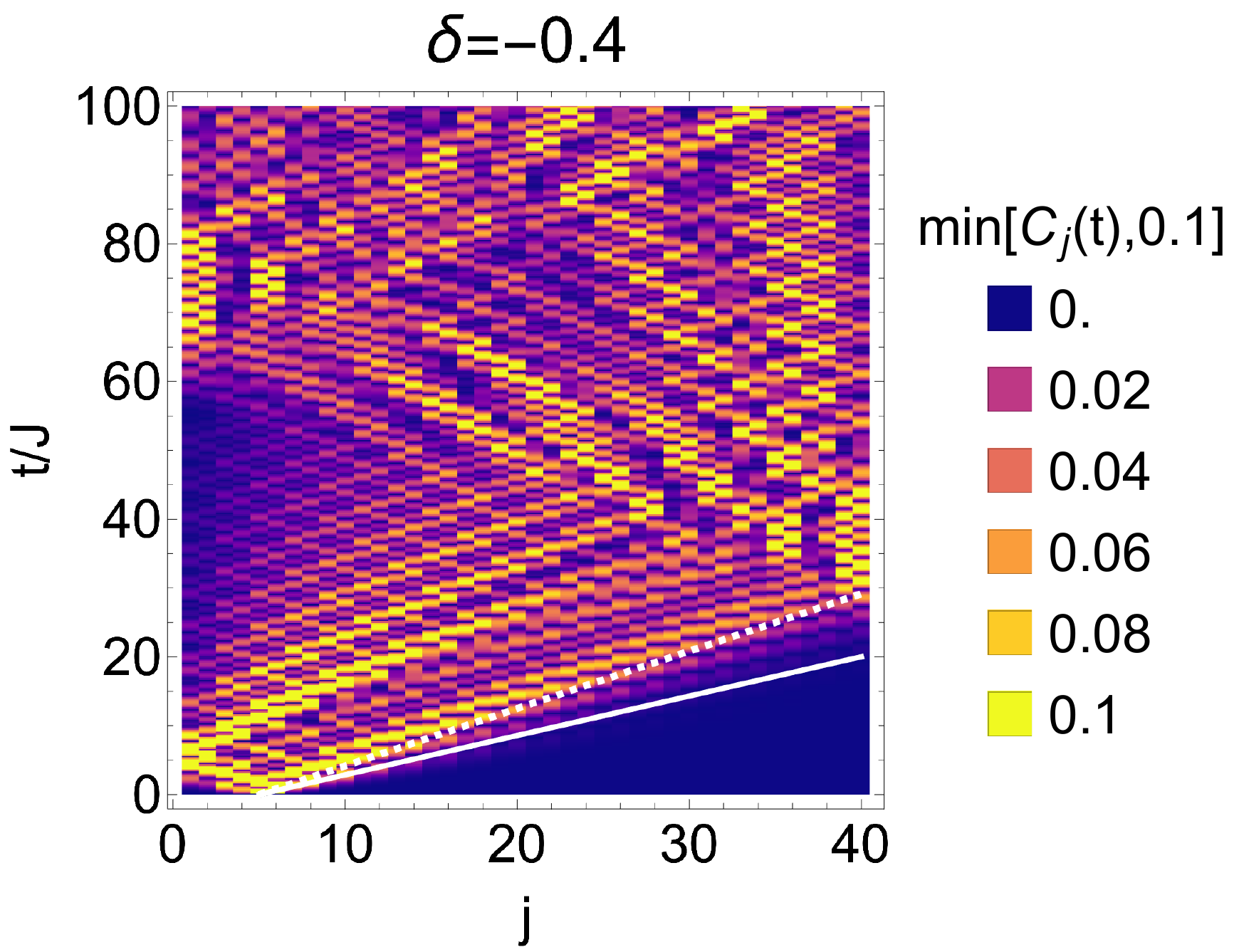}
	\caption{The OTOC $C_j(t)$ for the SSH model. The perturbation occurs at $j_0=5$ and the system size is $N=40$. The white lines are an aid to the eye for how fast the correlations spread, they show the maximum group velocity of the bulk bands $v_g=2J(1-|\delta|)$ (dashed lines) and the butterfly velocity $v_b$ (solid lines) found from fits to Eq.~\eqref{cfit}, see App.~\ref{app:fits}. The trapping of the information in the edge state can still be seen, even when the perturbation occurs further away from the edge. See also Fig.~\ref{figssh1}.
	\label{figssh2}}
\end{figure}

\begin{figure}
	\includegraphics[width=0.9\columnwidth]{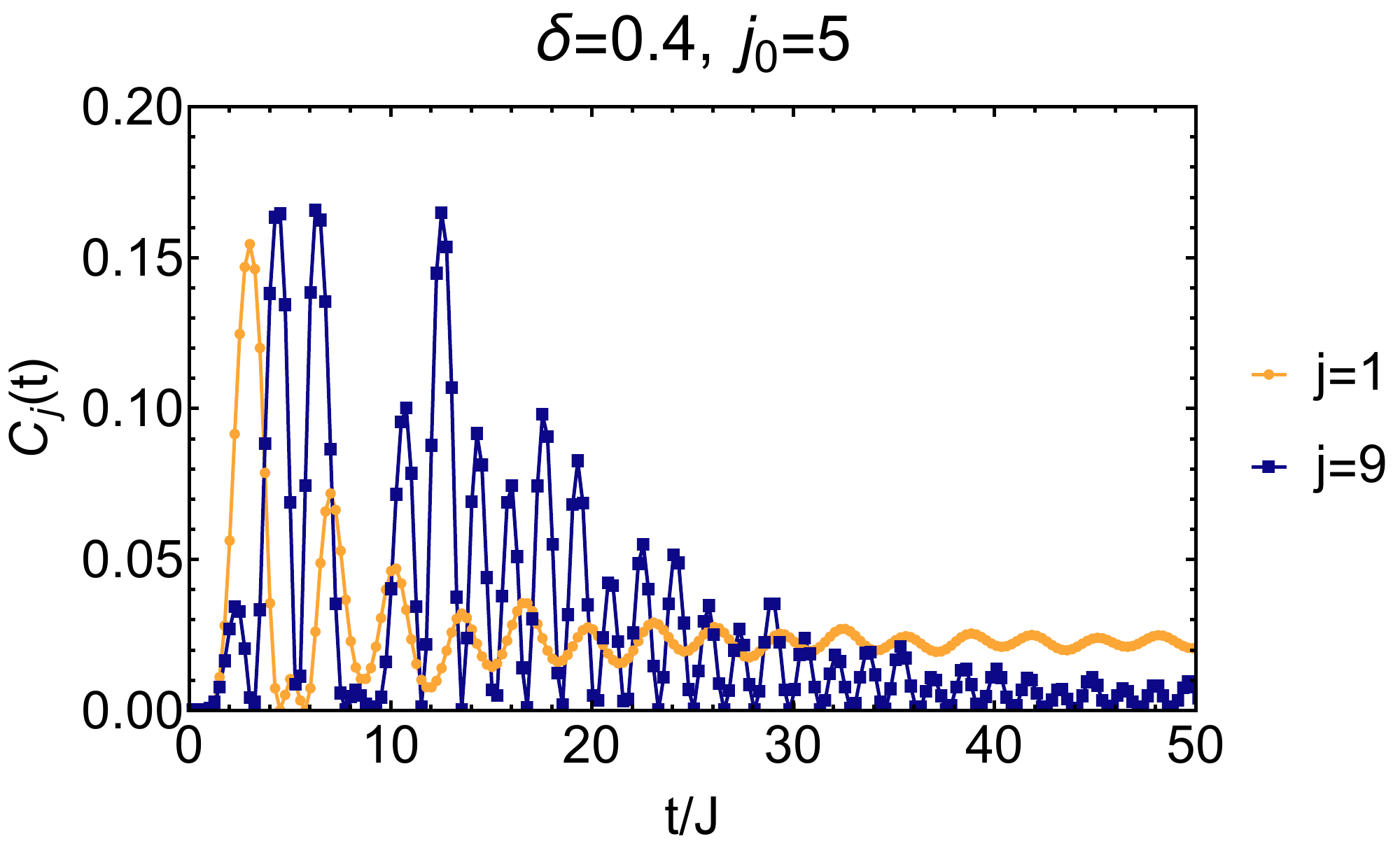}
	\caption{The OTOC $C_j(t)$ for the SSH model near the boundary. The perturbation occurs at $j_0=5$ and the system size is $N=40$. Information trapping in site $j=1$ is clearly visible. By contrast correlations in site, $j=9$ fade away after the transient behavior.}
	\label{figsshposlt}
\end{figure}

We can also perturb the chain further from the edge, see Fig.~\ref{figssh2}. The trapping of information in the topologically protected edge modes is still clearly visible, and does not occur when there is no edge state present. In principle one would expect trapping also on the opposite boundary after the correlations have reached there, however this effect is too small to be clearly visible in this plot. In Fig.~\ref{figsshposlt} we focus on two sites that demonstrate this trapping in the topologically non-trivial regime. We pick two sites at equal distances from the perturbation, one on the edge and one further inside the chain. Information becomes trapped inside the edge mode at site $j=1$, but naturally not further inside the chain where only bulk states have any weight.

\begin{figure}
	\includegraphics[width=0.9\columnwidth]{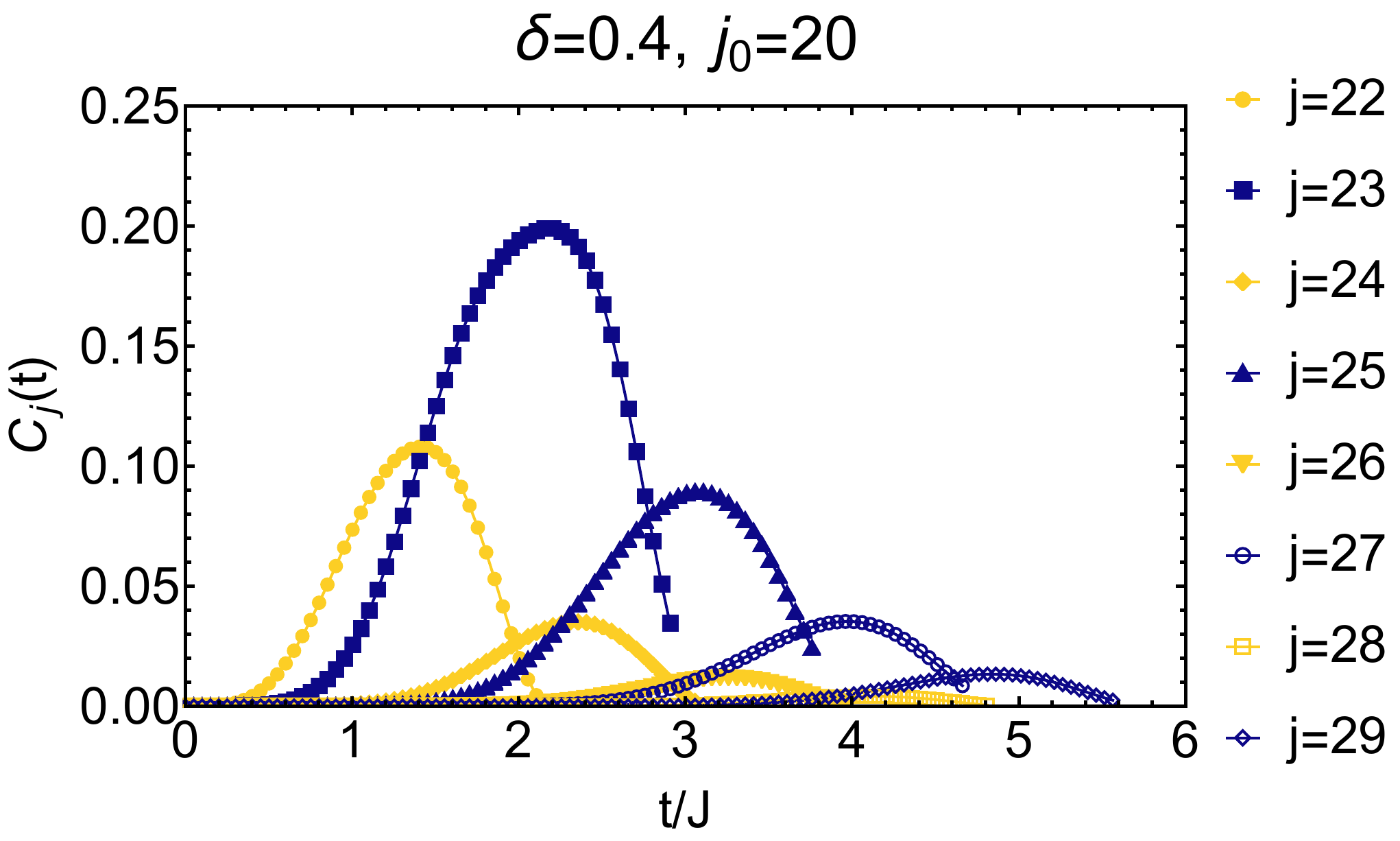}
	\caption{The OTOC $C_j(t)$ for the SSH model. The perturbation occurs at $j_0=20$ and the system size is $N=40$. Only the first onset of the correlation is shown to aid clarity.}
	\label{figsshpos2}
\end{figure}

As a result of the dimerized hopping of the SSH model, a clear even-odd effect in the perturbations can be seen. In Fig.~\ref{figsshpos2} we plot the OTOC for a series of distances from the perturbation. Here we focus on the bulk, and consider the non-trivial phase, though similar results can be seen for the trivial phase. The system is perturbed at $j=20$, and the hopping terms coupling even-odd sites are large in this case. For odd-even the hopping is small. It follows that $C_j(t)$ should be larger for odd $j$. This can be seen in Fig.~\ref{figsshpos2}, along with the observation that the peaks and onset of correlations are shifted for odd and even sites. We note that despite the local lack of reflection symmetry at any site where we perturb, there is no resultant chiral effect in the velocity~\cite{Sekania2021}, which is the same for left-moving and right-moving terms, due to these even-odd effects becoming averaged out.

%%%%%%%%%%%%%%%%%%%%%%%%%%%%%%%%%%%%%%%%%%%%%
\section{Results for the long range Kitaev model}\label{sec:lrk}
%%%%%%%%%%%%%%%%%%%%%%%%%%%%%%%%%%%%%%%%%%%%%

In this section, we turn to a more general one-dimensional topological system, a long-range Kitaev chain, which allows us to probe different topological phases and different perturbations. All together this allows for many different possibilities. Broadly speaking however the results follow qualitatively those of the SSH model reported in the preceding section. As for the SSH model we find that quenches do not play an important role, at least for the cases we considered, and we again focus here on cases where the initial state is the ground state of the Hamiltonian $\hat H$.

\begin{figure}
	\includegraphics[width=0.49\columnwidth]{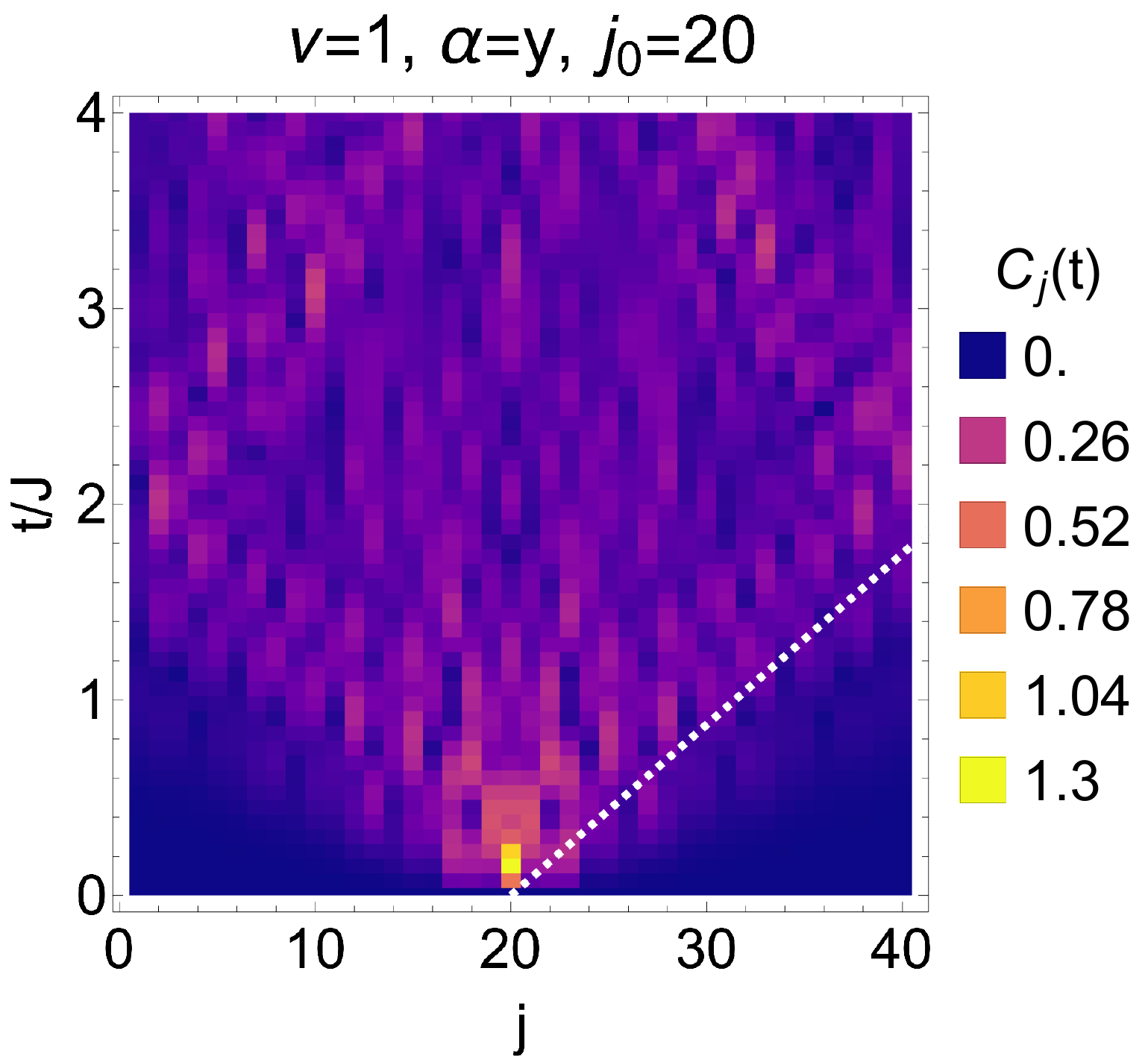}
	\includegraphics[width=0.49\columnwidth]{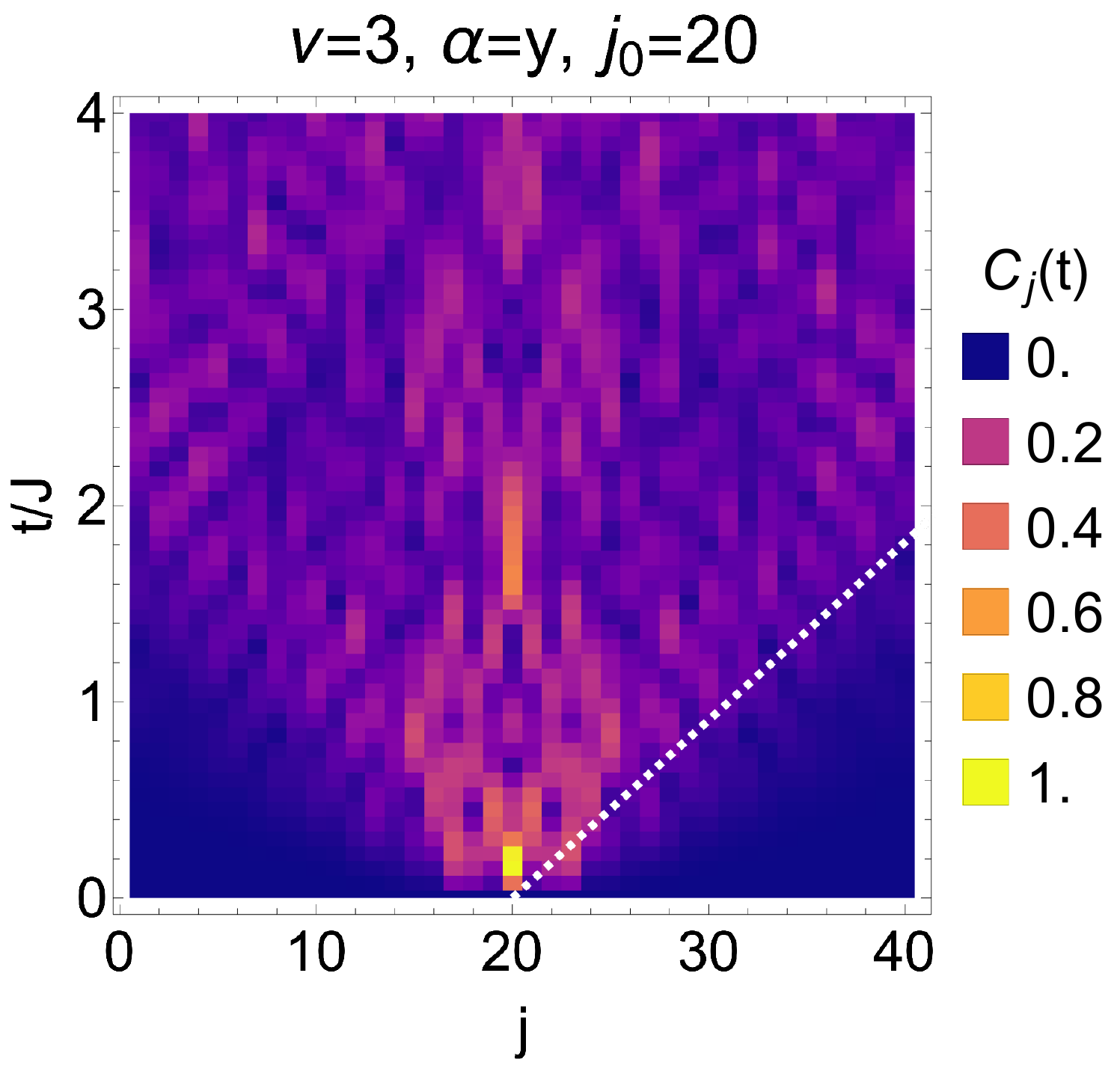}\\
	\includegraphics[width=0.49\columnwidth]{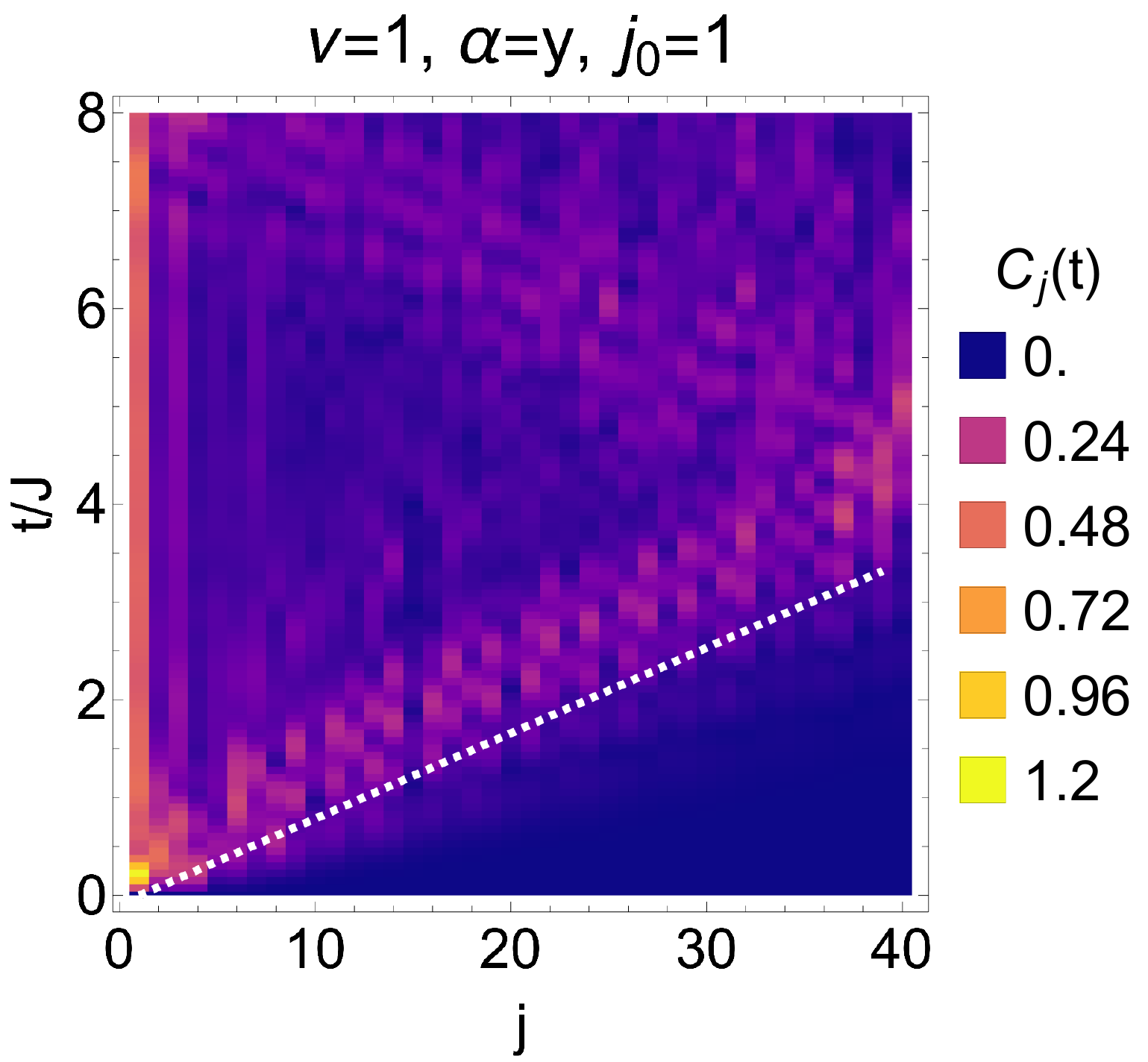}
	\includegraphics[width=0.49\columnwidth]{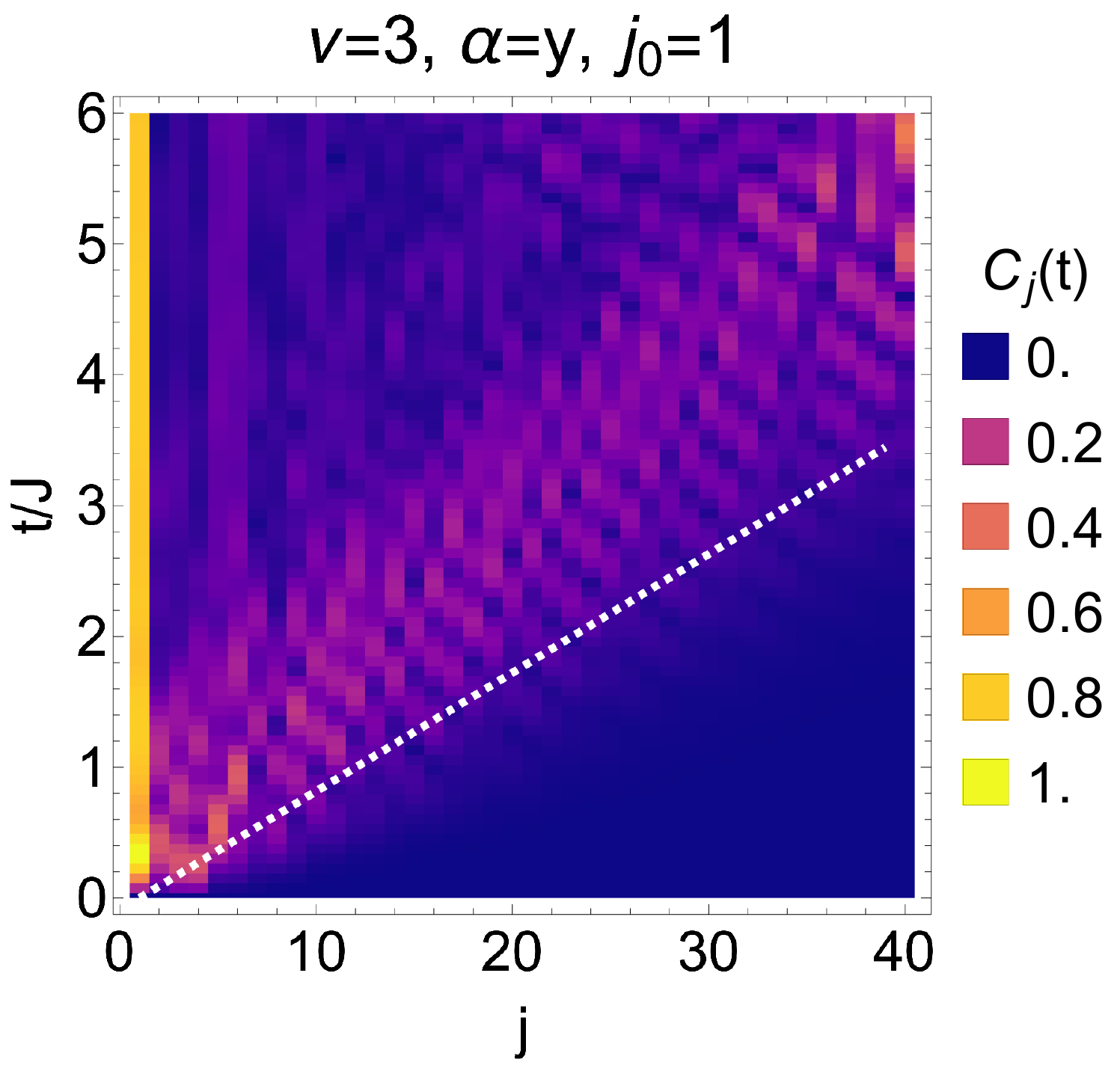}
	\caption{The OTOC $C_j(t)$ for the Kitaev model. The perturbation occurs at $j_0=20$ in the upper panels and $j_0=1$ in the lower panels and the system size is $N=40$. Results for $\hat H$ in two different topological phases are shown: $\nu=1$ and $\nu=3$. The dashed white lines are an aid to the eye for how fast the correlations spread, they show the maximum group velocity of the bulk bands. The perturbation is $\hat W_{j_0}=\exp\left(i W_0\hat \Psi^\dagger_{j_0}{\bm\tau}^y\hat \Psi_{j_0}\right)$.}
	\label{figlrkit}
\end{figure}

\begin{figure}
	\includegraphics[width=0.49\columnwidth]{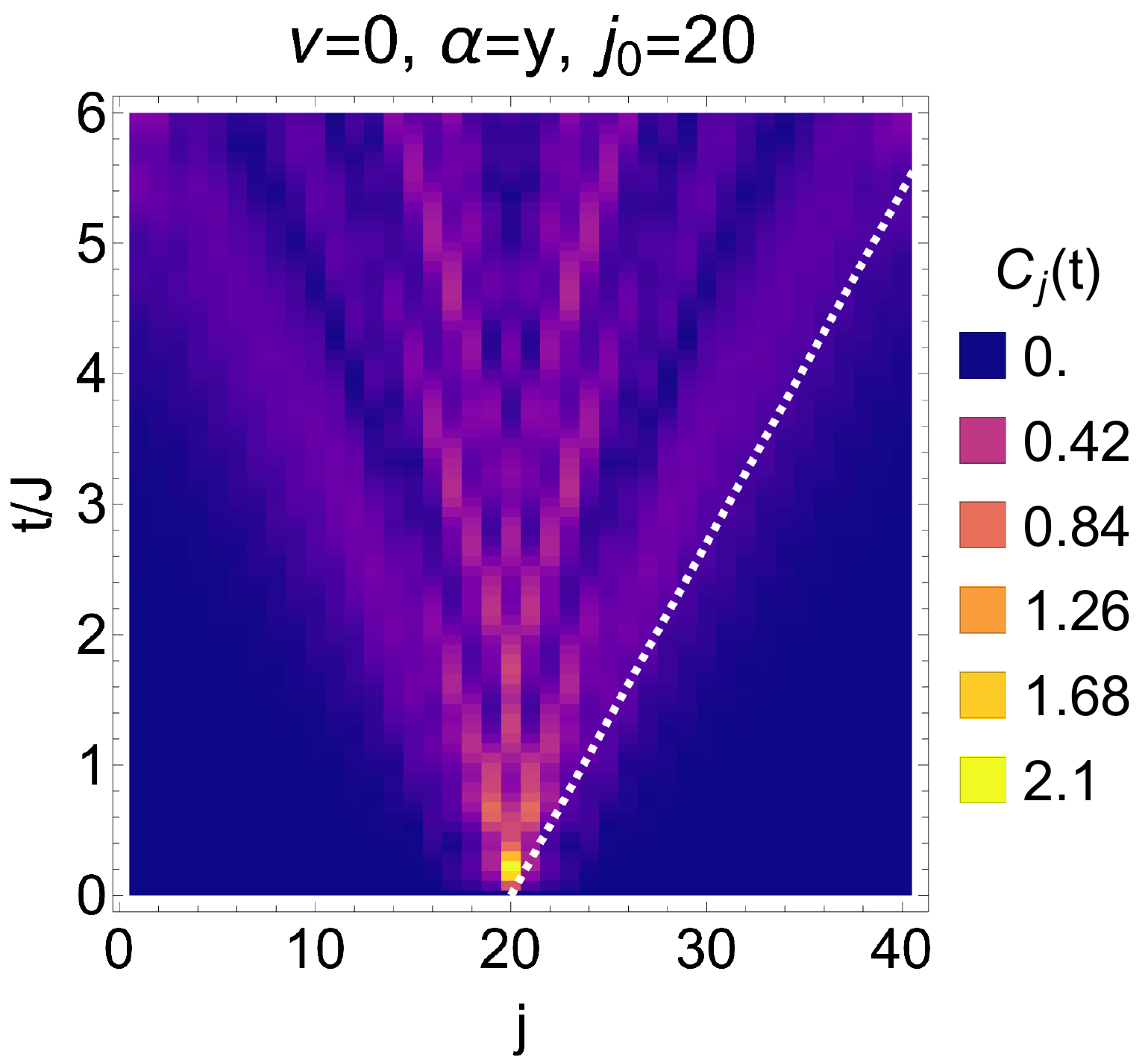}
	\includegraphics[width=0.49\columnwidth]{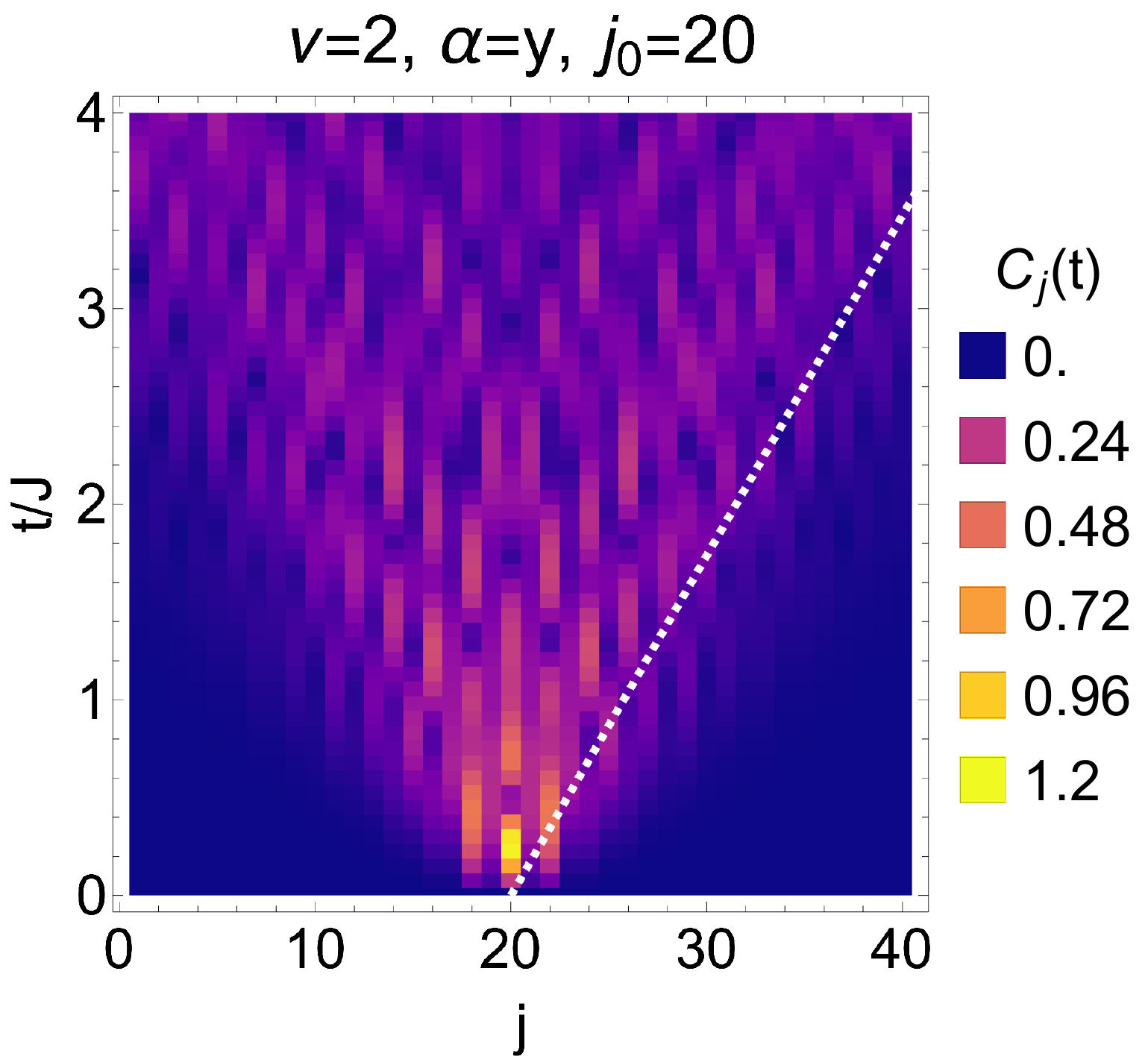}\\
	\includegraphics[width=0.49\columnwidth]{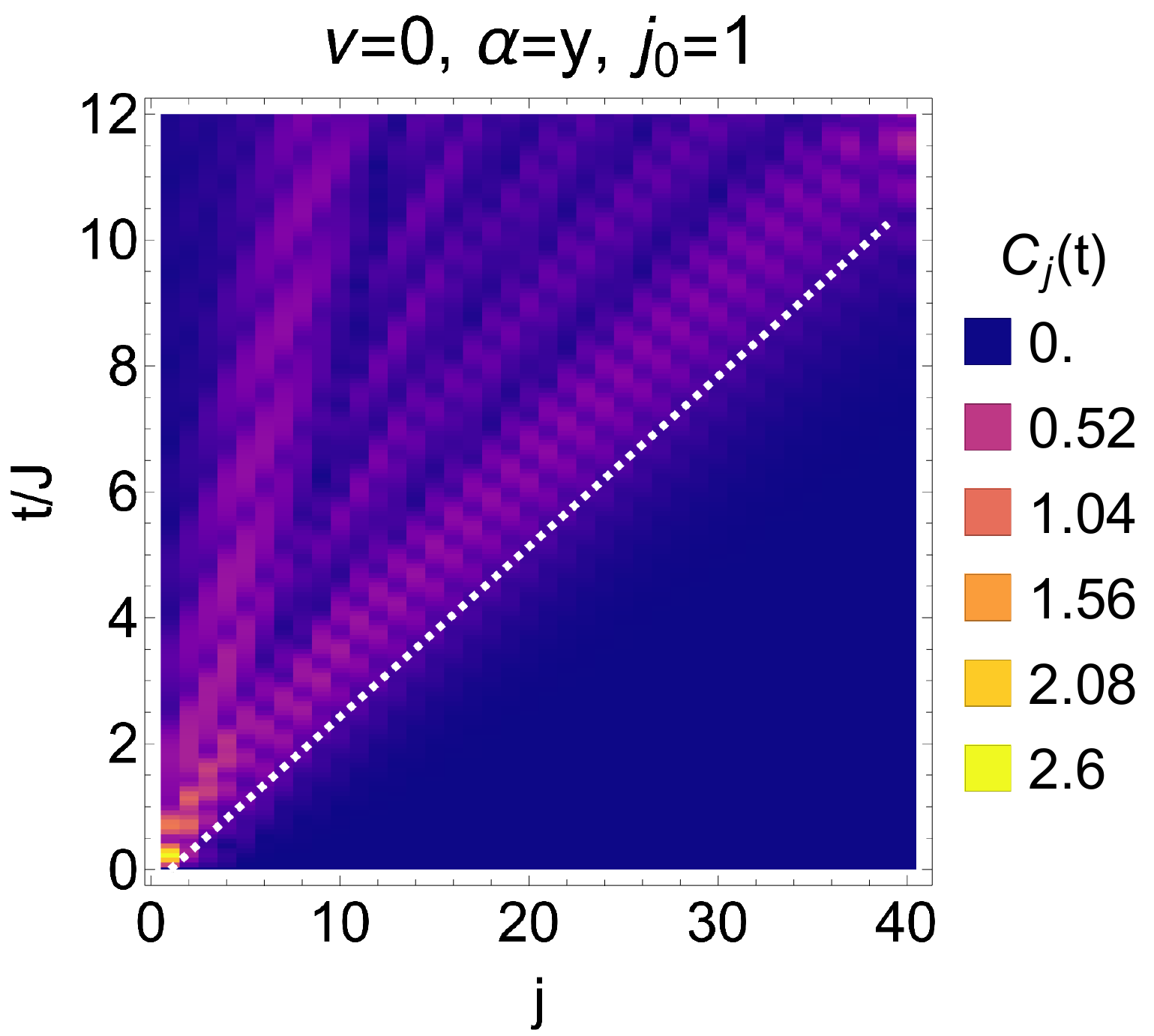}
	\includegraphics[width=0.49\columnwidth]{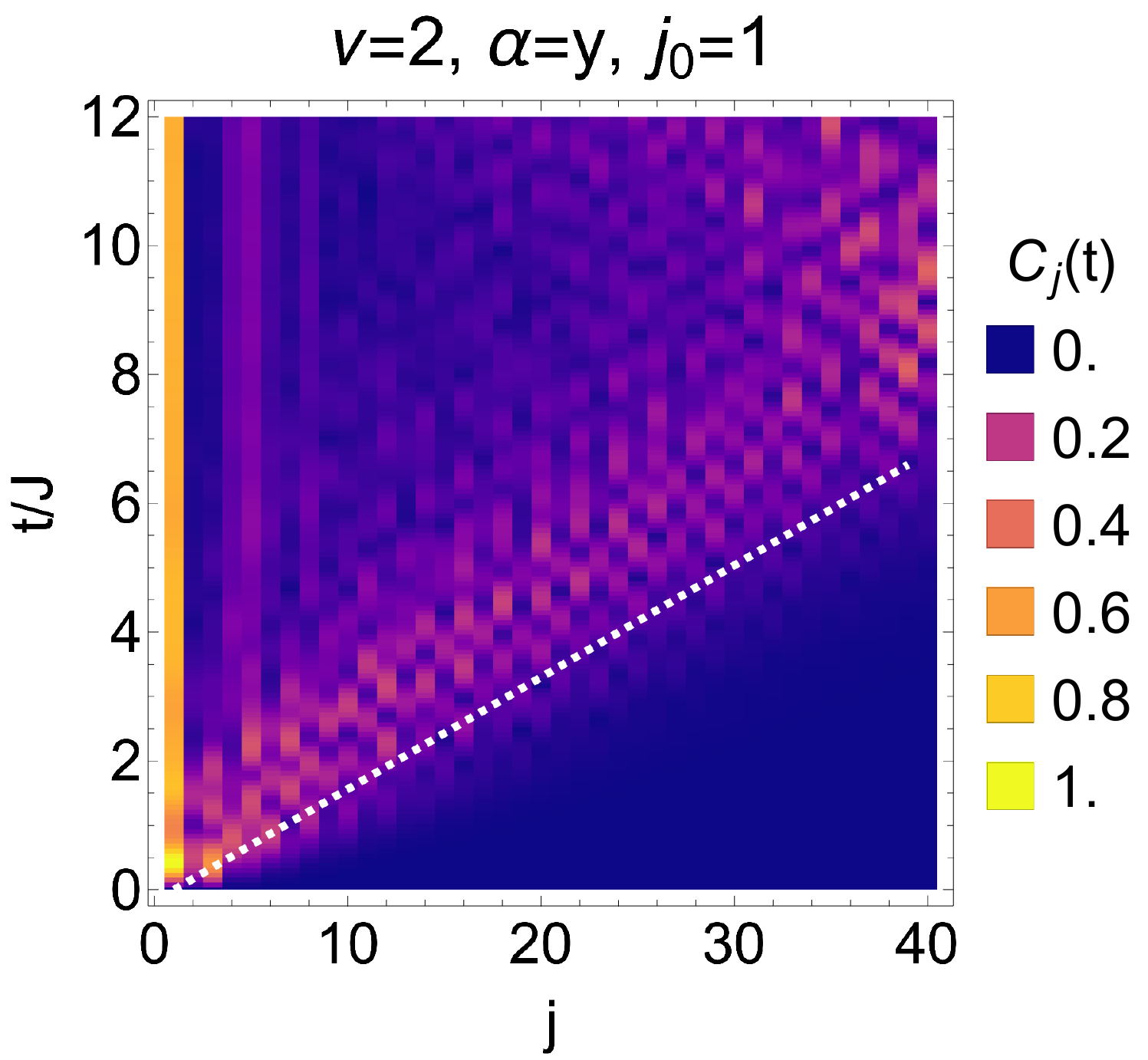}
	\caption{The OTOC $C_j(t)$ for the Kitaev model. The perturbation occurs at $j_0=20$ in the upper panels and $j_0=1$ in the lower panels and the system size is $N=40$. Results for $\hat H$ in two different topological phases are shown: $\nu=0$ and $\nu=2$. The dashed white lines are an aid to the eye for how fast the correlations spread, they show the maximum group velocity of the bulk bands. The perturbation is $\hat W_{j_0}=\exp\left(i W_0\hat \Psi^\dagger_{j_0}{\bm\tau}^y\hat \Psi_{j_0}\right)$.}
	\label{figlrkit2}
\end{figure}

In Figs.~\ref{figlrkit} and \ref{figlrkit2} we show exemplary results for perturbations $\hat W_{j_0}=e^{i W_0\hat \Psi^\dagger_{j_0}{\bm\tau}^y\hat \Psi_{j_0}}$ for $\hat H$ in topological phases with $\nu=0,1,2,3$. Actual parameters are given in Table \ref{tab:quenches}. Qualitatively speaking the scrambling demonstrated by $C_j(t)$ is similar in the bulk of all phases, though of course the butterfly velocity depends on details of the Hamiltonian. For the boundary we see, as for the SSH model, that the presence of topologically protected edge states traps information at the boundary.

\begin{figure}
	\includegraphics[width=0.9\columnwidth]{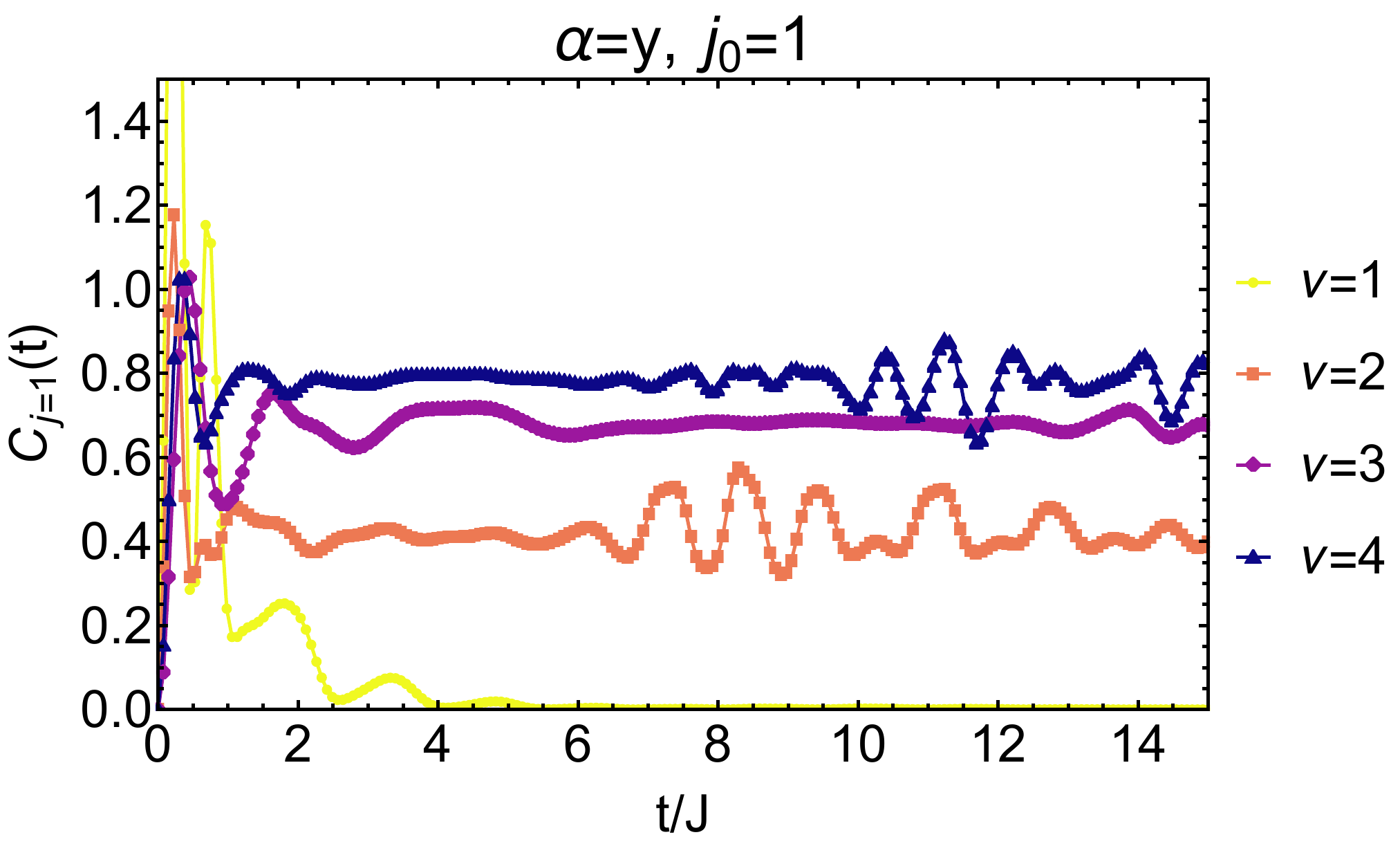}
	\caption{The OTOC $C_{j=1}(t)$ for the Kitaev model. The perturbation also occurs at $j_0=1$ and the system size is $N=40$. Results for $\hat H$ in all four possible different topological phases are shown. The perturbation is $\hat W_{j_0}=e^{i W_0\hat \Psi^\dagger_{j_0}{\bm\tau}^\alpha\hat \Psi_{j_0}}$ with here $\alpha=y$.}
	\label{figlrkitedge}
\end{figure}

In Fig.~\ref{figlrkitedge} we compare the information trapped at the boundary for the different topological phases. This appears to correlate with the topological  invariant, i.e.~with the number of boundary modes. Here we show results for $\alpha=y$, similar results are seen for $\alpha=z$. However for a perturbation $\alpha=0$ or $\alpha=x$ the direction of the trend is reversed. Of course, in all cases, no information is trapped for the $\nu=0$ phase.

We have also considered a series of other possible perturbations, some additional results are shown in App.~\ref{app:extrakit}. In principle, we can vary not only the type of perturbation, but we can have different perturbations $\hat W_{j_0}=\exp\left(i W_0\hat \Psi^\dagger_{j_0}{\bm\tau}^\alpha\hat \Psi_{j_0}\right)$ and $\hat V_j=\exp\left(i V_0\hat \Psi^\dagger_j{\bm\tau}^\beta\hat \Psi_j\right)$ with $\alpha\neq\beta$. We find that our results look qualitatively the same for all combinations of $\alpha$ and $\beta$. It also seems to make no difference whether the perturbation breaks the symmetries of the symmetry-protected topological phase. 

%%%%%%%%%%%%%%%%%%%%%%%%%%%%%%%%%%%%%%%%%%%%%
\section{Analytical results}\label{sec:an}
%%%%%%%%%%%%%%%%%%%%%%%%%%%%%%%%%%%%%%%%%%%%%

The models focused on here also allow for some analytical expressions to be found. Here we focus on the SSH model, which has a semi-analytical solution also for open boundary conditions~\cite{Shin1997,Sirker2014}. We note however that the results do not give a numerical advantage over the formalism used in the previous sections. First let us note that we can rewrite the unitary perturbations as
\begin{equation}
	\hat W_{j_0}=1+\hat n_{j_0}\left(e^{iW_0}-1\right),
\end{equation}
and
\begin{equation}
	\hat V_{j}=1+\hat n_{j}\left(e^{iV_0}-1\right).
\end{equation}
We find therefore that we can rewrite
\begin{align}
    C_j(t)&=\left\langle\left|\left[\hat W_{j_o}(t),\hat V_{j}\right]\right|^2\right\rangle\\\nonumber
    &=4\underbrace{\left(1-\cos W_0\right)\left(1-\cos V_0\right)}_{\equiv \mathcal{A}(V_0,W_0)}\left\langle\left|\left[\hat n_{j_o}(t),\hat n_{j}\right]\right|^2\right\rangle.
\end{align}
Furthermore $\hat n_{j_0}(t)$ can be found by transforming to the eigenbasis. To do this let us focus on the case where both $j$ and $j_0$ are even sites, so that $\hat n_{j_0}=\hat a_{j_0}^\dagger\hat a_{j_0}$ in the notation of Eq.~\eqref{RM_model}. By Fourier transforming and defining the annihilation operators of the eigenstates as
\begin{align}
	\hat\alpha_k&=\frac{1}{\sqrt{2}}\left(A_k\hat a_{k}+\hat b_{k}\right)\textrm{ and}\nonumber\\
	\hat\beta_k&=\frac{1}{\sqrt{2}}\left(-A_k\hat a_{k}+\hat b_{k}\right),
\end{align}
the Hamiltonian becomes
\begin{equation}
	\hat H=\sum_k\epsilon_k\left(\hat\beta_k^\dagger\hat\beta_k-\hat\alpha_k^\dagger\hat\alpha_k\right)
\end{equation}
with $\epsilon_k=2J\sqrt{\cos^2k+\delta^2\sin^2k}$ and
\begin{equation}
	A_k=\frac{2J\left(\cos k+i\delta \sin k\right)}{\epsilon_k}.
\end{equation}
We find for the time evolved density
\begin{align}
	\hat n_{j_0}(t)=\frac{1}{N}\sum_{k,k'}\frac{e^{i(k-k')j_0}}{2A^*_kA_{k'}}
 &\left(e^{-i\epsilon_kt}\hat\alpha^\dagger_k-e^{i\epsilon_kt}\hat\beta^\dagger_k\right)\\\nonumber
 &\quad\times\left(e^{i\epsilon_{k'}t}\hat\alpha_{k'}-e^{-i\epsilon_{k'}t}\hat\beta_{k'}\right),
\end{align}
and a similar expression for $\hat n_j=\hat a^\dagger_j\hat a_j$, but with $t=0$ and $j_0\to j$. We note that all sums over momenta are of the form $k=2\pi n/N$ with $n=1,2,\ldots N/2$. 

Calculating the now trivial commutator and the expectation value over the half filled ground state we find
\begin{equation}\label{can}
	C_j(t)=2\mathcal{A}(V_0,W_0)
    \left[\left|c^0_j(t)\right|^2+c^1_j(t)-c^2_j(t)\right].
\end{equation}
The terms $c^i_j(t)$ are given by
\begin{equation}
	c^0_j(t)=\frac{1}{N^2}\sum_{q,q'}\sin\left[2\Delta j(q-q')-\epsilon_q t\right]
	\cos\left[\epsilon_{q'} t\right],
\end{equation}
where $\Delta j=j-j_0$,
\begin{equation}
	c^1_j(t)=
	\frac{1}{N^2}\sum_{q,q'}\cos\left[\epsilon_q t\right]\cos\left[\epsilon_{q'} t\right]\cos\left[2\Delta j(q-q')\right],
\end{equation}
and finally
\begin{align}
	c^2_j(t)=&
	\frac{1}{N^4}\sum_{k,k',q,q'}\cos\left[\epsilon_q t\right]\cos\left[\epsilon_{q'} t\right]\\\nonumber
    &\times\cos\left[(\epsilon_{k}-\epsilon_{k'})t-2\Delta j(k+k'-q-q')\right].
\end{align}

\begin{figure}
	\includegraphics[width=0.9\columnwidth]{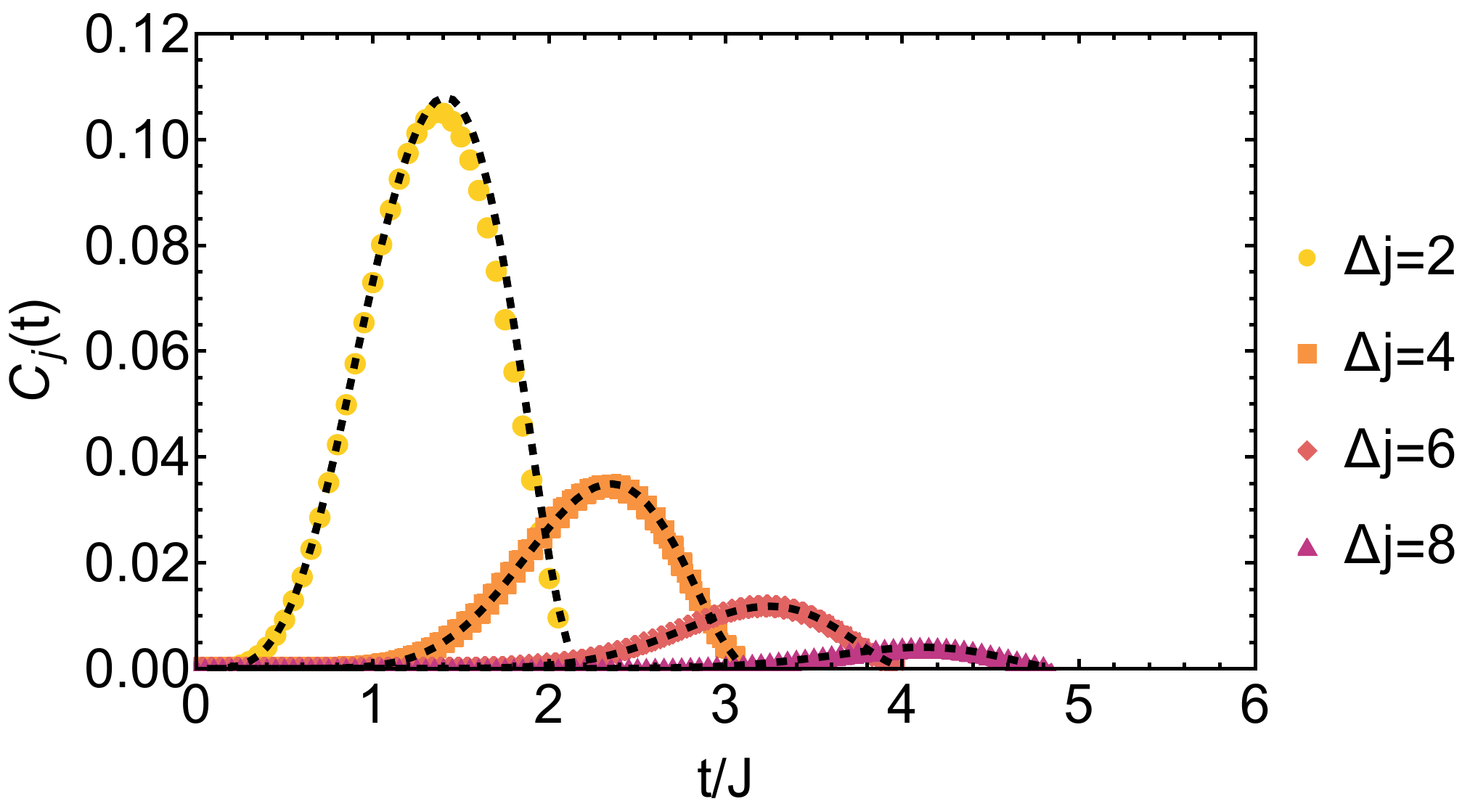}
	\caption{A comparison of the OTOC $C_j(t)$ in the bulk, for different distances $\Delta j$ and for the SSH model, found from Eq.~\eqref{eq:f} and Eq.~\eqref{can}. The former is for open boundary conditions, shown as dashed lines, the latter is a semi-analytical result for a chain with periodic boundary conditions, shown as coloured symbols. The system size is $N=40$ and results for the topologically non-trivial phase are shown. The agreement is good though we note that numerically Eq.~\eqref{eq:f} has the advantage for the cases we considered.}
	\label{figsshan1}
\end{figure}

\begin{figure}
	\includegraphics[width=0.9\columnwidth]{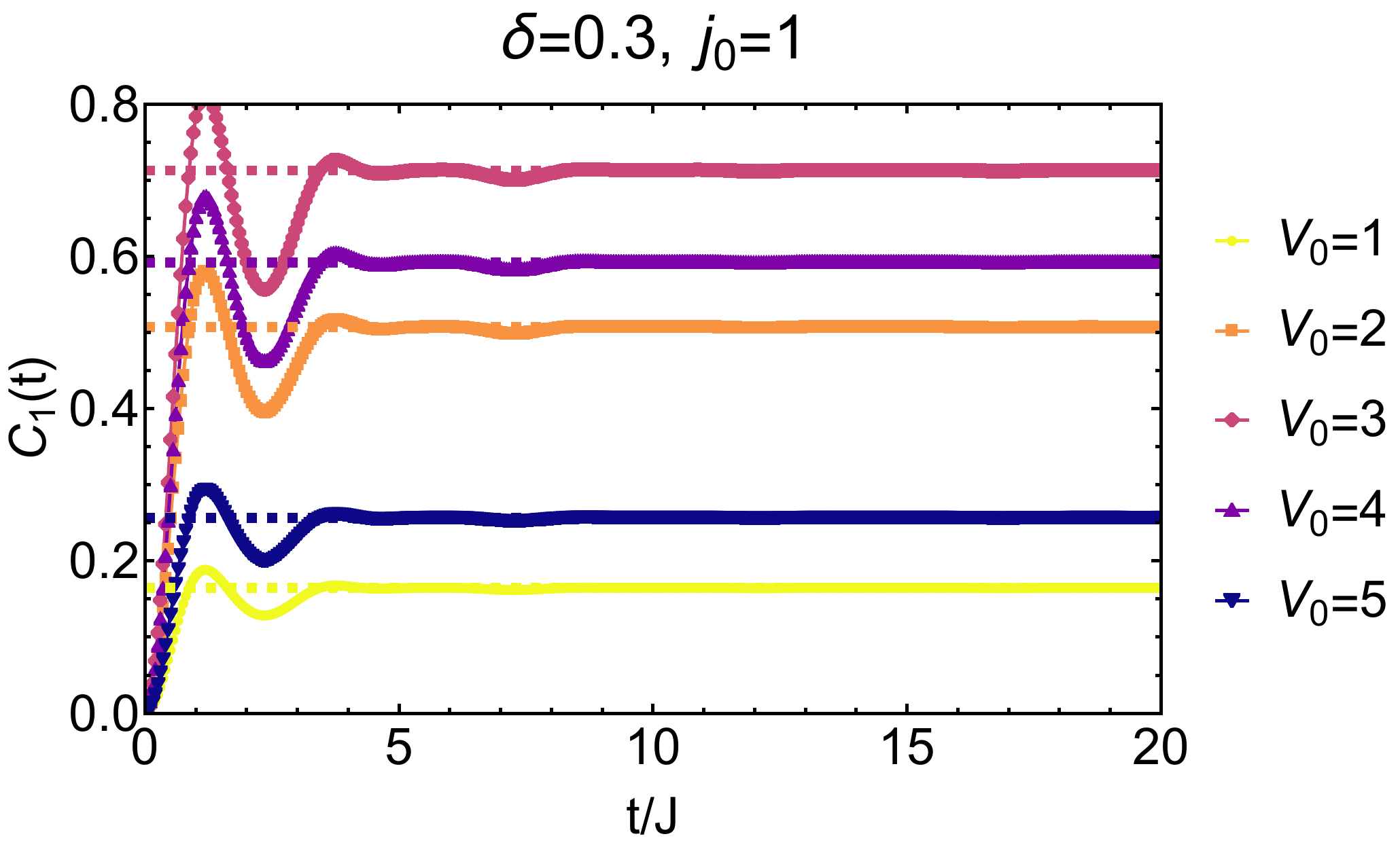}\\
	\includegraphics[width=0.9\columnwidth]{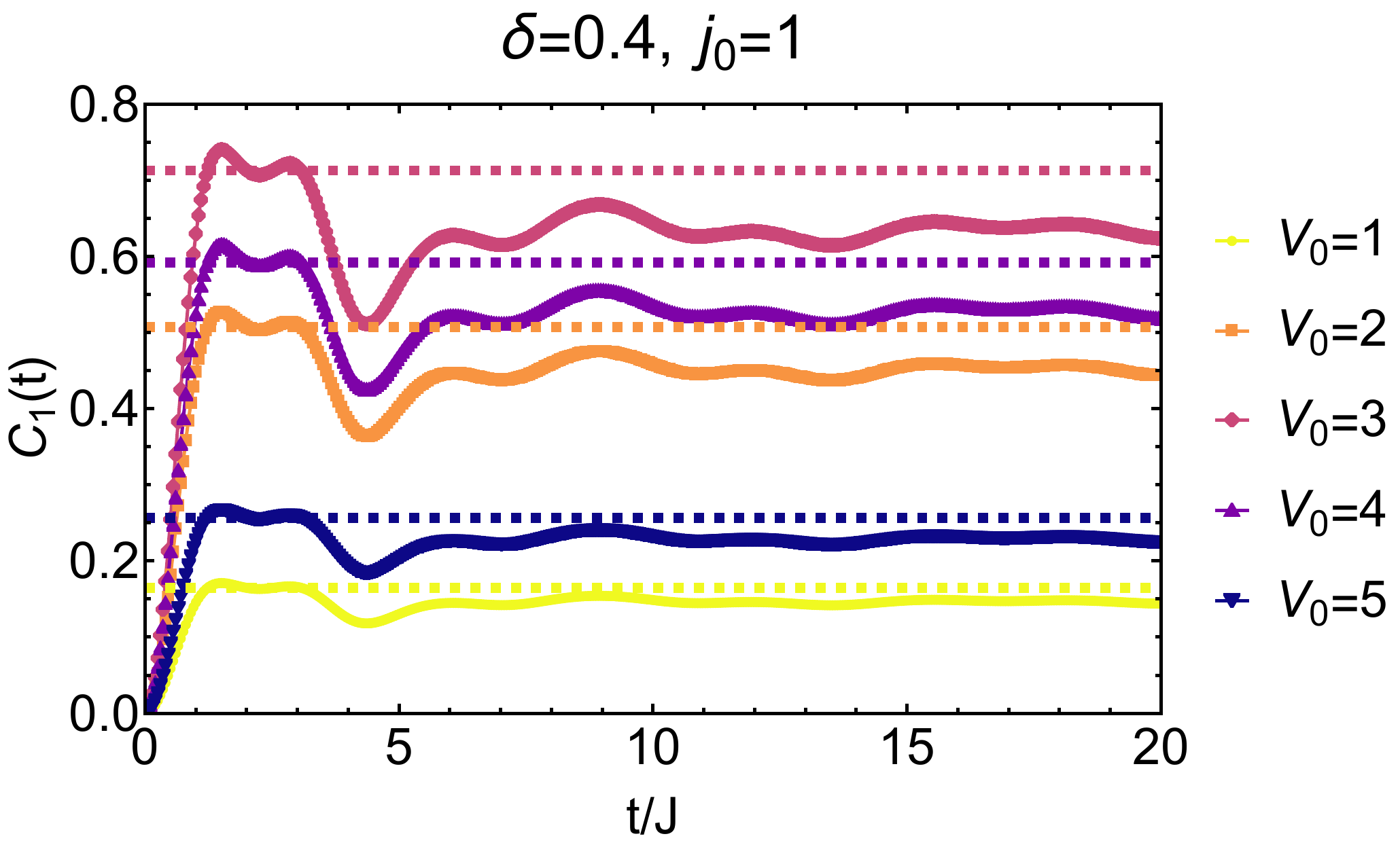}
	\caption{The OTOC $C_j(t)$ for the SSH model on the boundary for different dimerization strengths and perturbation strengths $V_0$, in all cases $W_0=5$. The perturbation occurs at $j_0=1$ and the system size is $N=40$. Results for the topologically non-trivial phase are shown. Dashed lines show $\mathcal{A}(V_0,W_0)/2=(1-\cos V_0)(1-\cos W_0)/2$, which is independent of $\delta$}
	\label{figsshan2}
\end{figure}

We can compare these expression directly to those found in Sec.~\ref{sec:ssh}, see Fig.~\ref{figsshan1}. For larger system sizes these calculations quickly become numerically slower than the formalism used in Sec.~\ref{sec:ssh}, even for $N=40$. Analysing these expressions reveals several details. First the strength of the perturbations used is purely a prefactor, this can also be seen in Fig.~\ref{figsshan2} where we vary $V_0$. Second the expressions depend only on the eigenenergies, the distance $\Delta j$, which states are filled, and naturally on time $t$.

If we focus now on the edge of the SSH system for open boundary conditions we can perform a similar calculation to determine the origin of the information trapping in the topologically non-trivial phases. We can again transform the density operators to the appropriate eigenbasis. In this case we can write that $\hat c_j=u_{jk}\hat \alpha_k$ with $k=1,2,\ldots N$. The eigenenergies are given by~\cite{Shin1997,Sirker2014}
\begin{equation}
    \lambda_k=\pm J\sqrt{2}\sqrt{1+\delta^2+(1-\delta^2)\cos\theta_k}
\end{equation}
with $\theta_k$ determined by
\begin{equation}
    (1-\delta)\frac{\sin\left[(N/2+1)\theta_k\right]}{\sin\theta_k}+(1+\delta)\frac{\sin\left[(N/2)\theta_k\right]}{\sin\theta_k}=0.
\end{equation}
In the topologically non-trivial regime complex $\theta_k$ solutions exist which give rise to the exponentially small ``zero energy modes'', one of which is filled in the ground state.

Focusing on the left edge and taking $j=j_0=1$ one finds
\begin{equation}
    C_1(t)=\frac{\mathcal{A}(V_0,W_0)}{N^2}\sum_{\substack{k\\\lambda_k>0}}\sum_{\substack{q\\\lambda_q<0}}\left|\left[e^{-i\lambda_qt}\mathcal{F}_t-e^{i\lambda_kt}\mathcal{F}^*_t\right]\right|^2,
\end{equation}
where 
\begin{equation}
    \mathcal{F}_t=\frac{1}{N}\sum_pe^{i\lambda_pt}.
\end{equation}
In the topologically non-trivial regime it can be seen that there is a static contribution from the zero mode to $\mathcal{F}_t$ resulting in a static term in $C_1(t)$ of the form
\begin{equation}
    \frac{2\mathcal{A}(V_0,W_0)}{N^4}\sum_{\substack{k\\\lambda_k>0}}\sum_{\substack{q\\\lambda_q<0}}\sum_{p,p'}=\frac{\mathcal{A}(V_0,W_0)}{2}.
\end{equation}
Interestingly this is completely independent of the dimerization strength, though this is only the simplest case for which the perturbations occur on the same site.

In Fig.~\ref{figsshan2} we compare the dynamics on the edge for two different dimerization strengths and for different perturbation strengths $V_0$. For smaller dimerization strengths we find that the dynamics of $C_1(t)$ are closely tied to the static contribution for long times, for larger dimerizations the corrections are larger, but the trend is still apparent. We can therefore see that the pinning in the edge nodes is largely due to the localized zero energy modes, which due to having exponentially small energies have little intrinsic dynamics.

%%%%%%%%%%%%%%%%%%%%%%%%%%%%%%%%%%%%%%%%%%%%%
\section{Conclusions}\label{sec:concl}
%%%%%%%%%%%%%%%%%%%%%%%%%%%%%%%%%%%%%%%%%%%%%

In this article, we have considered the role that topologically protected edge states have in scrambling in topological matter. We find that when taking the OTOC average over a typical ground state, the topological phase of this ground state does not play a significant role. The time evolving Hamiltonian is the key factor. We find that bulk scrambling of information does not significantly depend on the topological phase, and from fits to the data we could extract the butterfly velocity of the spread of the correlations. For the SSH chain, a clear odd-even effect depending on the dimerization was visible. The nature of the perturbation also did not play a role within the limits we tried, and we considered several different local unitary perturbations, including those which break the symmetries protecting the topological order.

The boundaries do appear in some cases to accelerate the spread of correlations, possibly from  scattering from the boundary. The main finding of this article is that information becomes trapped in the edge modes of the system, an effect completely absent when $\hat H$ is topologically trivial. By calculating an expression for the OTOC analytically we find that the information trapping is caused by the very slow dynamics of the edge modes present in the topologically non-trivial phase due to their exponentially small energies. Once this mode is perturbed there is a static contribution to the OTOC, which we refer to as information trapping. Furthermore, when considering phases with larger invariants, and therefore larger numbers of boundary modes, there appears to be a correlation between the number of boundary modes and the magnitude of $C_j(t)$ trapped. This occurs most clearly for perturbations that occur at or near the edge modes, but is still visible once correlations spread to the localized edge modes from even further away.

We propose that this may be used as a signature of the edge modes which could be applied in various out-0of-equilibrium scenarios, including local quenches and boundary distortions. Extensions to two dimensional systems with either helical or chiral edge modes would be an interesting extension of this work.

%%%%%%%%%%%%%%%%%%%%%%%%%%%%%%%%%%%%%
\acknowledgments
This work was supported by the National Science Centre (NCN, Poland) under the grant 2019/35/B/ST3/03625 (HC and NS). Data can be found on Zenodo~\cite{SedlmayrZ}.
%%%%%%%%%%%%%%%%%%%%%%%%%%%%%%%%%%%%%

\appendix

\section{OTOCs following quenches}\label{app:quenches}

\begin{figure}
	\includegraphics[width=0.49\columnwidth]{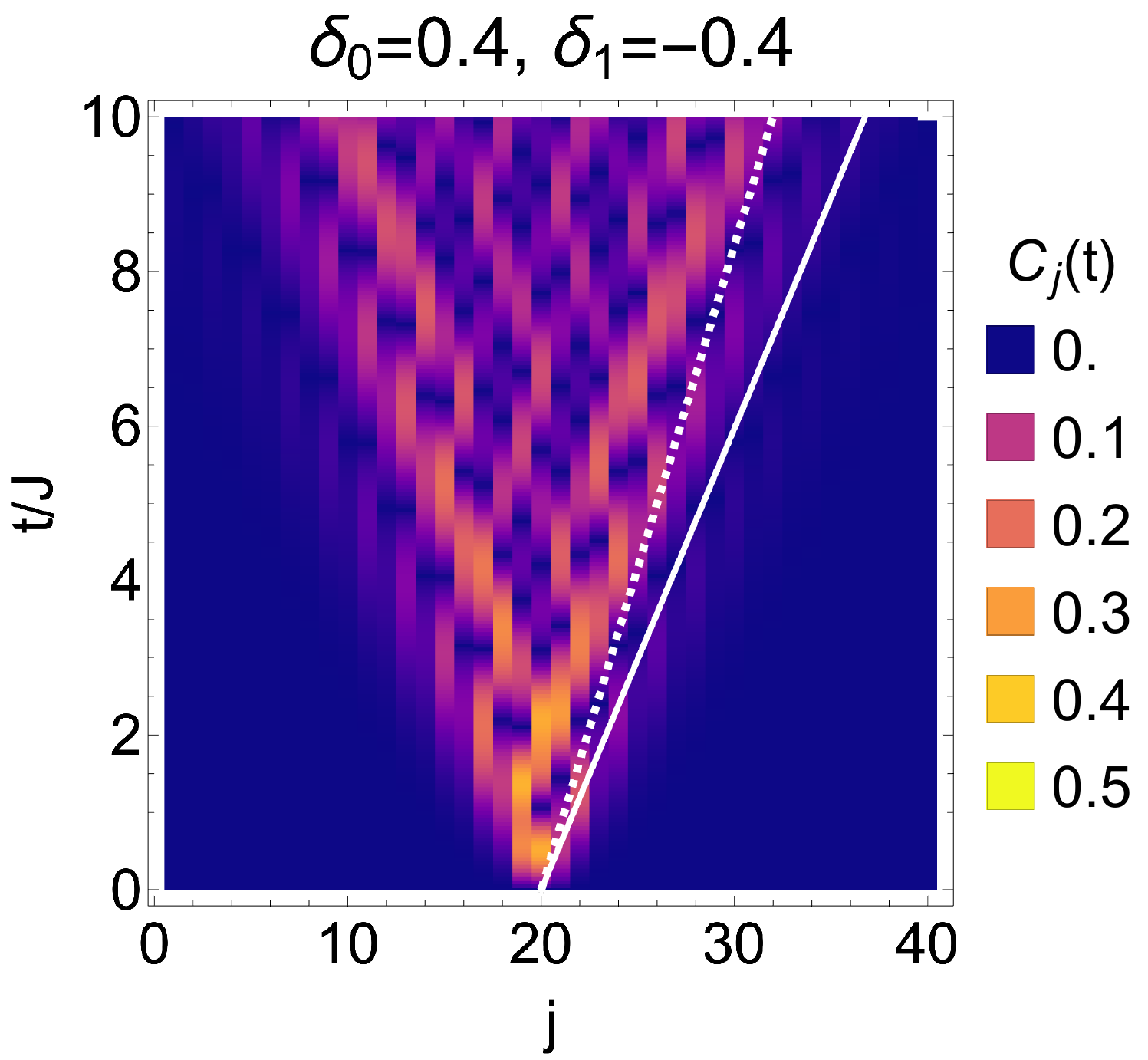}
	\includegraphics[width=0.49\columnwidth]{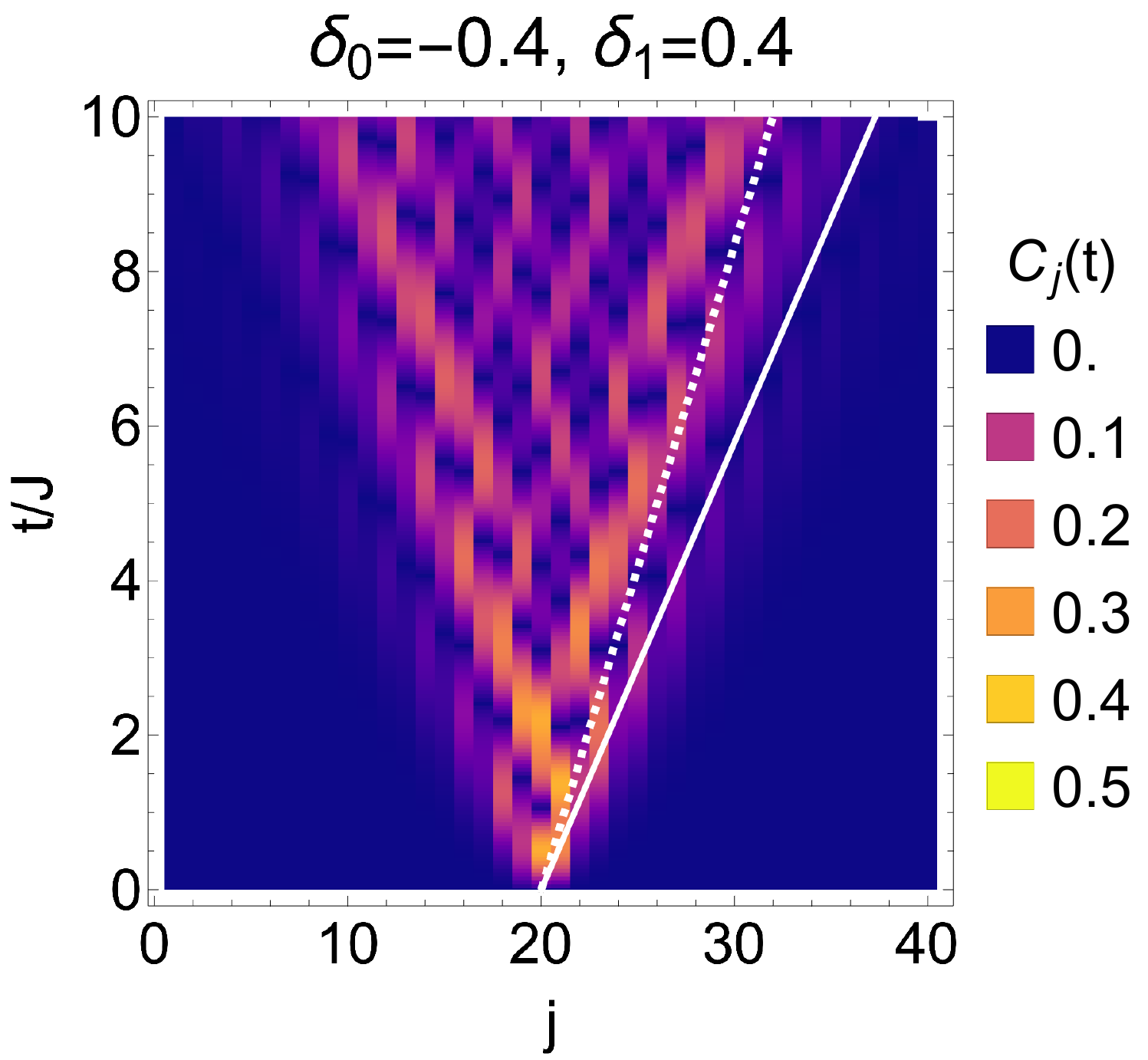}\\
	\includegraphics[width=0.49\columnwidth]{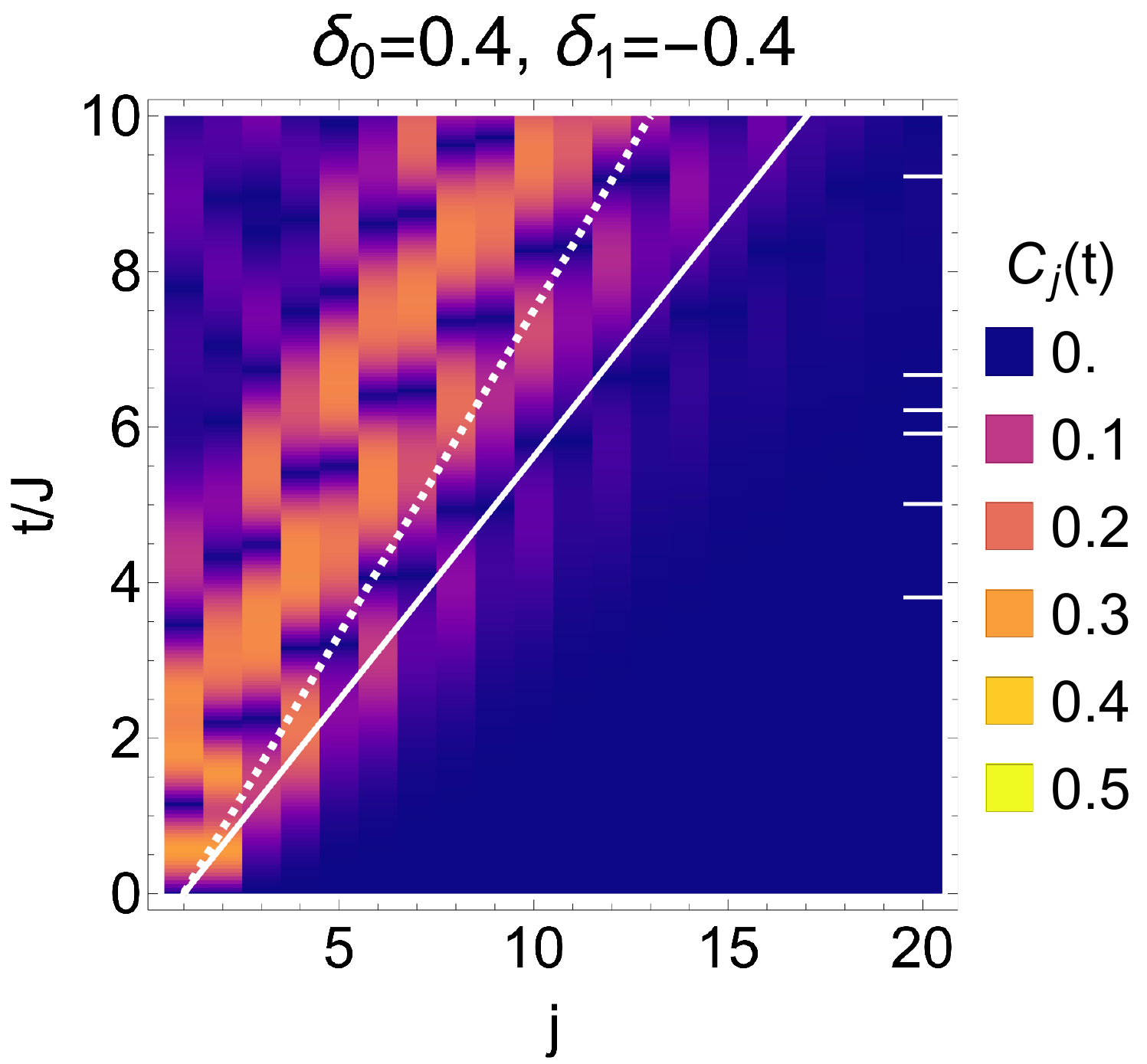}
	\includegraphics[width=0.49\columnwidth]{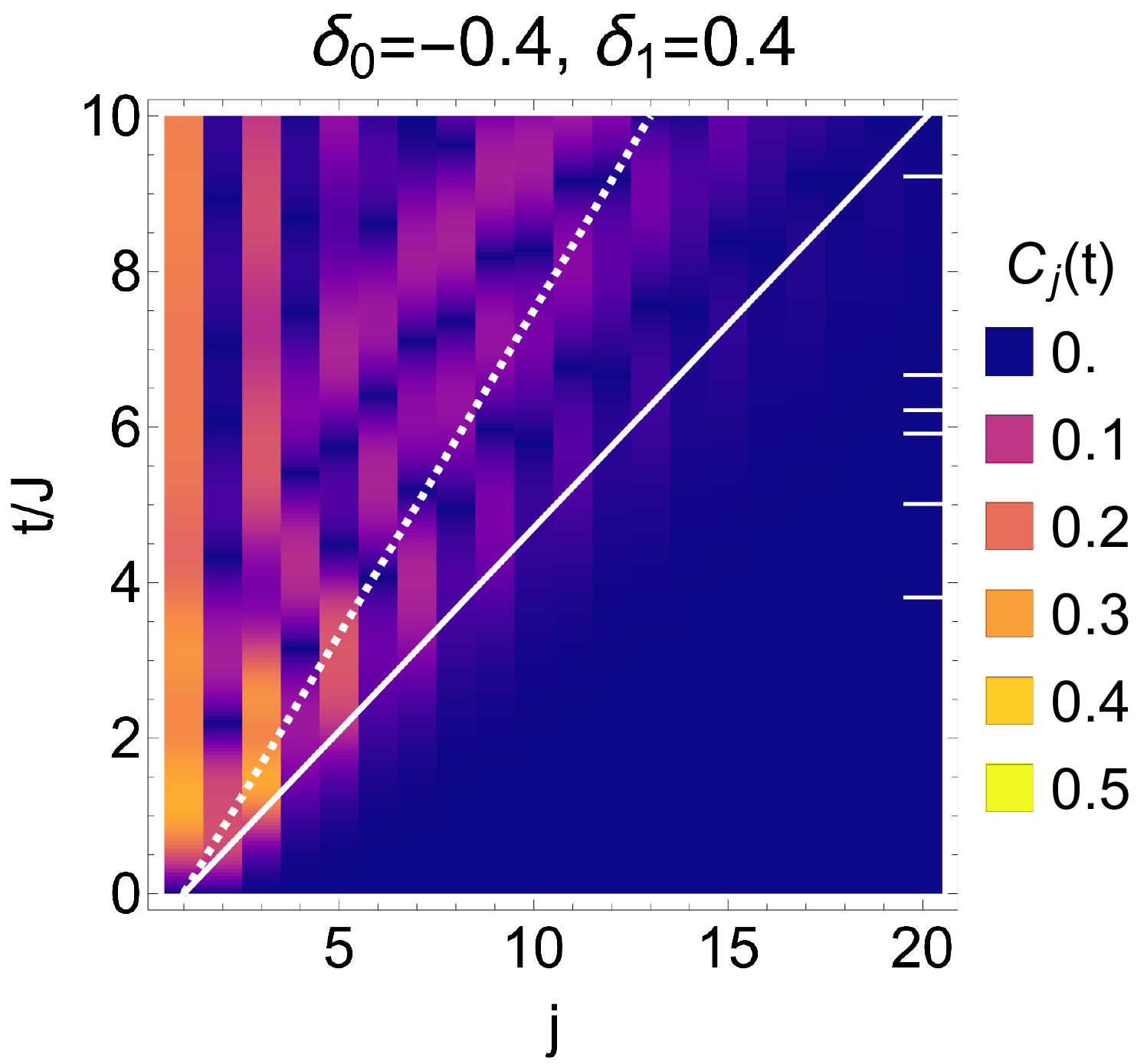}
	\caption{The OTOC $C_j(t)$ for the SSH model when the initial state is the ground state of $\hat H(\delta_0)$ and time evolution occurs for $\hat H(\delta_1)$. The perturbation occurs at $j_0=20$ in the upper panels and $j_0=1$ in the lower panels and the system size is $N=40$. The white lines are an aid to the eye for how fast the correlations spread, they show the maximum group velocity of the bulk bands $v_g=2J(1-|\delta|)$ (dashed lines) and the butterfly velocity $v_b$ (solid lines) found from fits to Eq.~\eqref{cfit}, see App.~\ref{app:fits}.}
	\label{appfig1}
\end{figure}

It is a natural question to ask if it makes any significant difference to the results whether the initial state of the system is chosen to be a state other than an eigenstate of $\hat H$. It is well known that for topological phases quenches that cross phase boundaries lead to dynamical quantum phase transitions~\cite{Vajna2015}, the definition of which is in terms of the Loschmidt echo, a closely related concept to OTOCs. Furthermore, this leads to  a dynamical bulk boundary correspondence~\cite{Sedlmayr2018} which depends on the direction of the quench. Here we find no dependence on the nature of the initial state or whether a quench is performed (within reason). However, similarly to the dynamical bulk boundary correspondence~\cite{Maslowski2020}, it is only the time-evolving Hamiltonian $\hat H$ which is crucial for the effects we see, such as information trapping at the boundary. See Fig.~\ref{appfig1} for explicit examples.

\section{Fits for Lieb-Robinson bounds and the Lyapunov exponent}\label{app:fits}

\begin{figure}[!ht]
	\includegraphics[width=0.8\columnwidth]{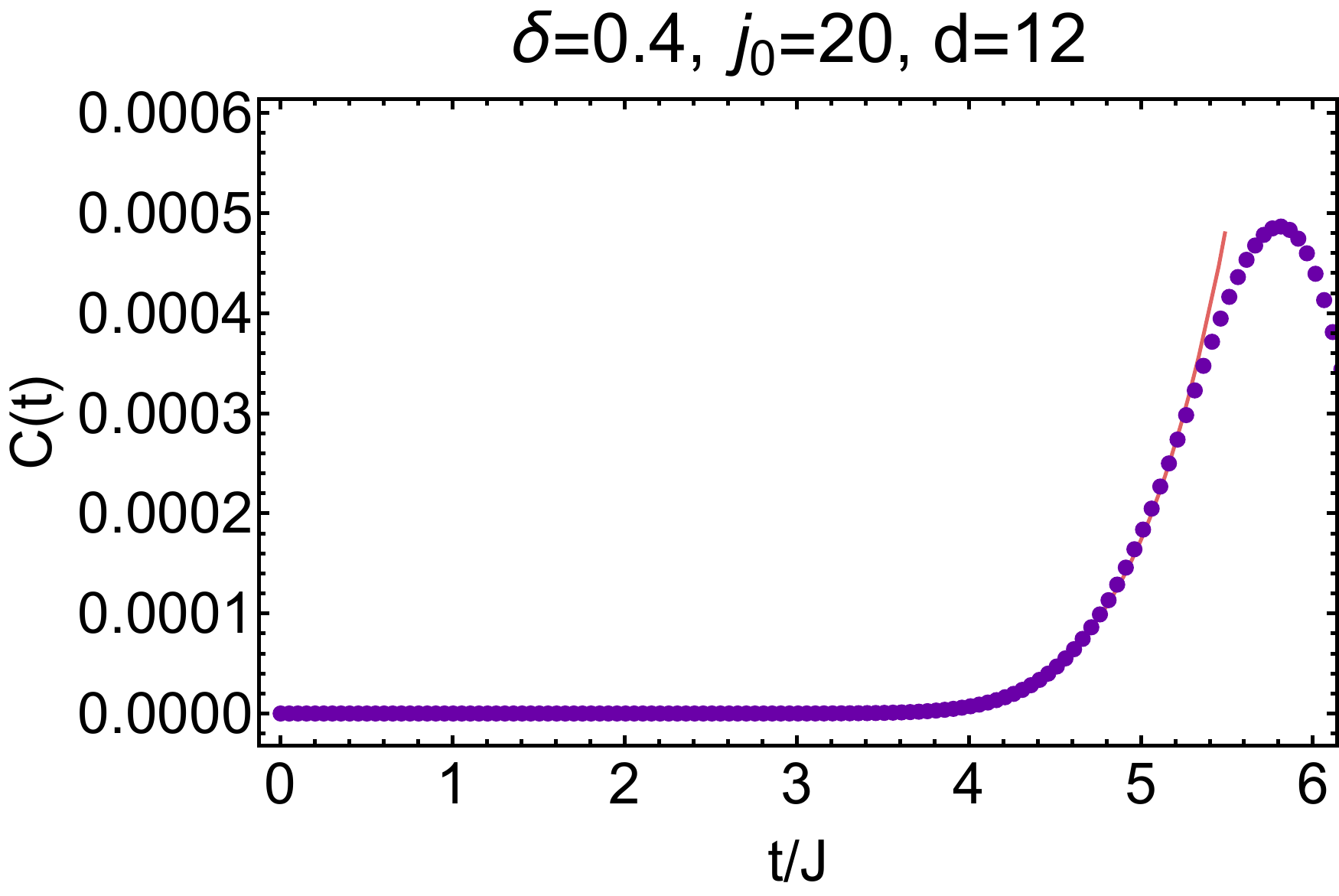}
	\caption{An example of the fitting for the SSH model: $C_j(t)\sim e^{\lambda_L\left(t-d/v_b\right)^\frac{2}{3}/t^\frac{1}{2}}$ where $d$ is the distance from the perturbation to site $j$, $\lambda_L$ is the Lyapunov exponent and $v_b$ the butterfly velocity. $\lambda_L$, $v_b$, and the overall amplitude are the fitting parameters.}
	\label{figapp2}
\end{figure}

For very short times we can find a reasonable fit between Eq.~\eqref{cfit} and the numerical data. However it should be stressed that this fit is for very short times only, but as seen in the main text does give reasonable bounds for the butterfly velocity compared to the numerical results. Examples of the fits are given in Fig.~\ref{figapp2}. For the Kitaev chain we could not find reasonable fits. We note that our model is not of course a truly long-ranged model, possessing only hopping terms that extend to the third nearest neighbour sites.

\section{Different perturbations for the Kitaev chain}
\label{app:extrakit}

\begin{figure}
	\includegraphics[width=0.49\columnwidth]{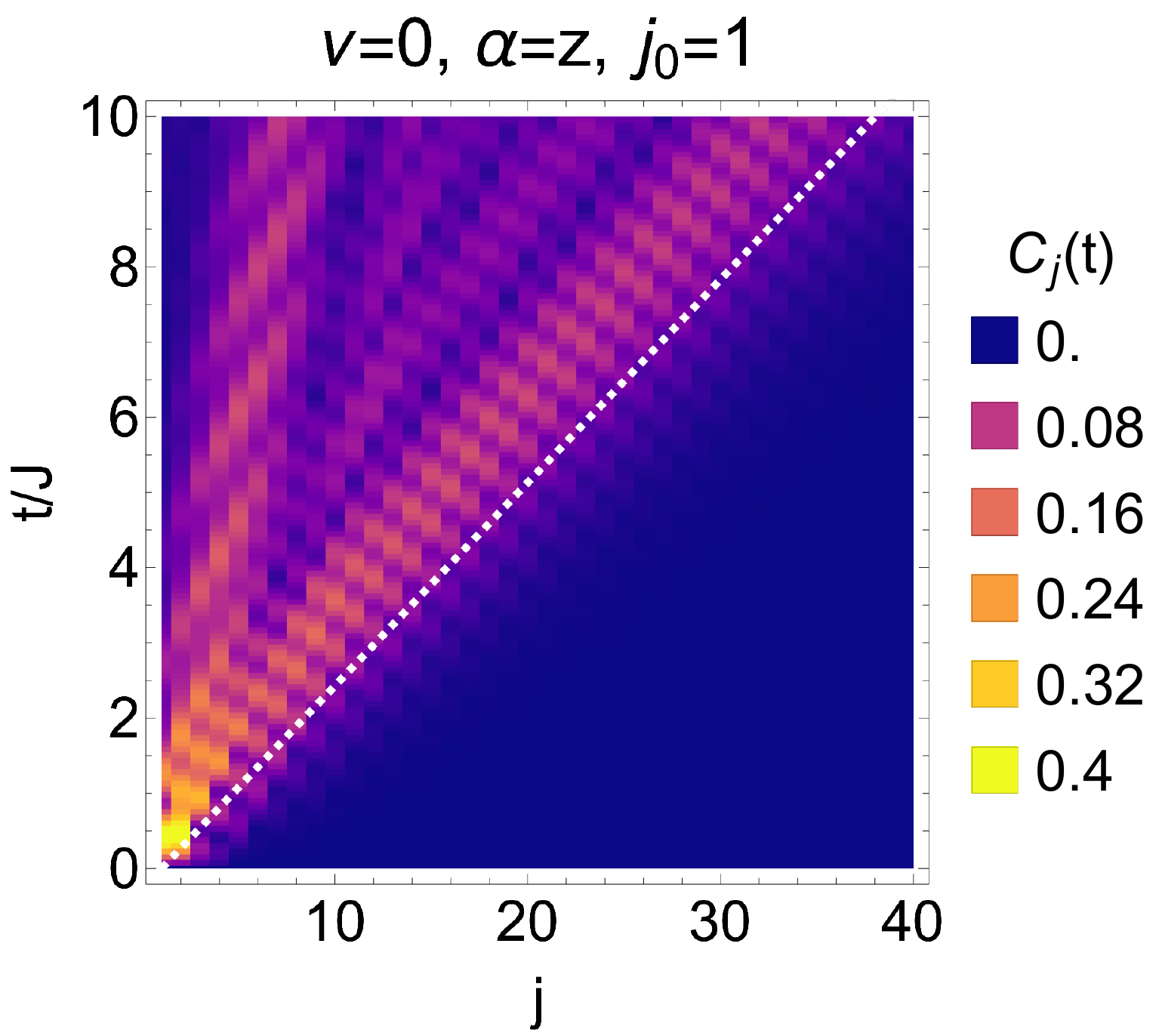}
	\includegraphics[width=0.49\columnwidth]{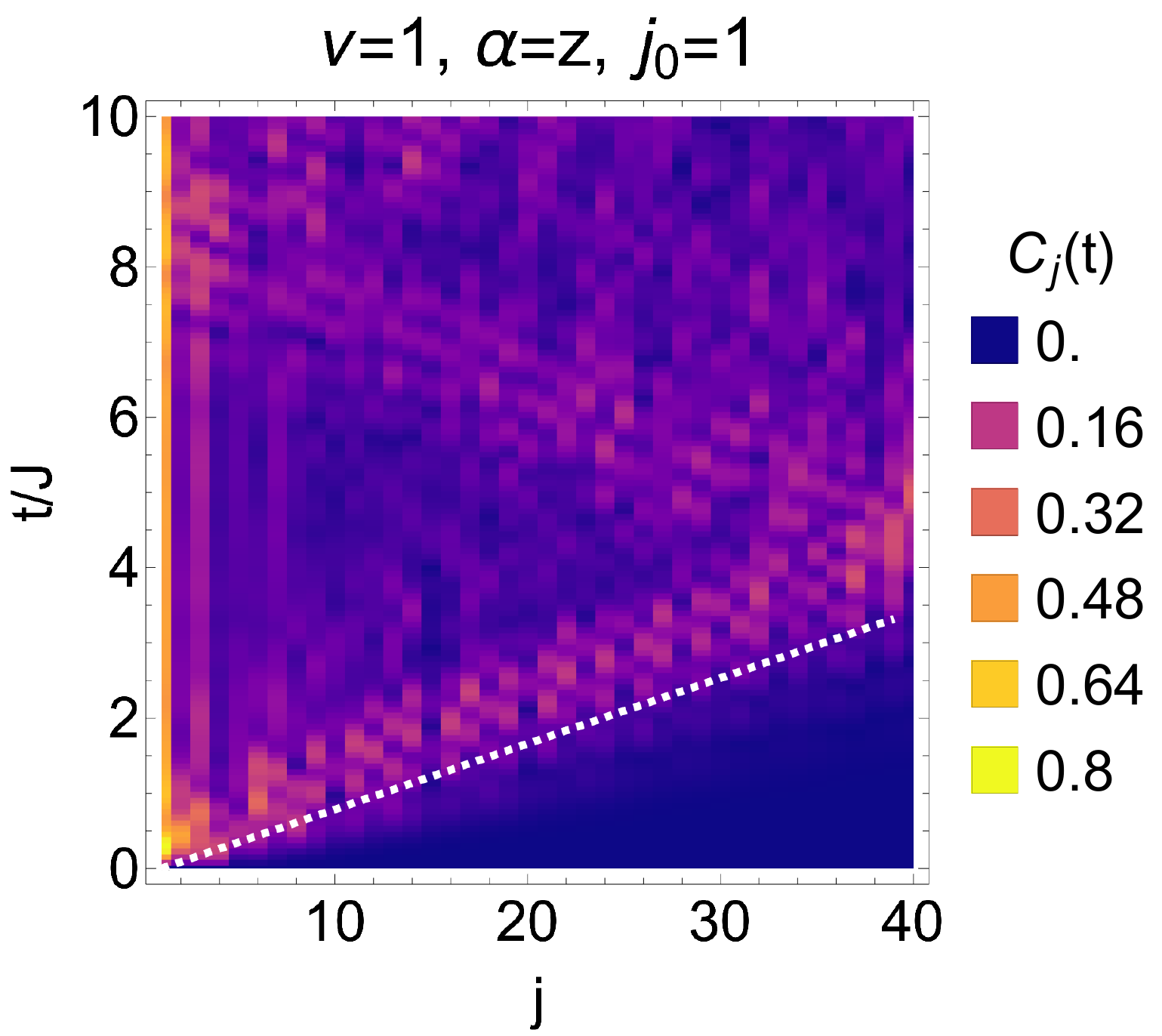}\\
	\includegraphics[width=0.49\columnwidth]{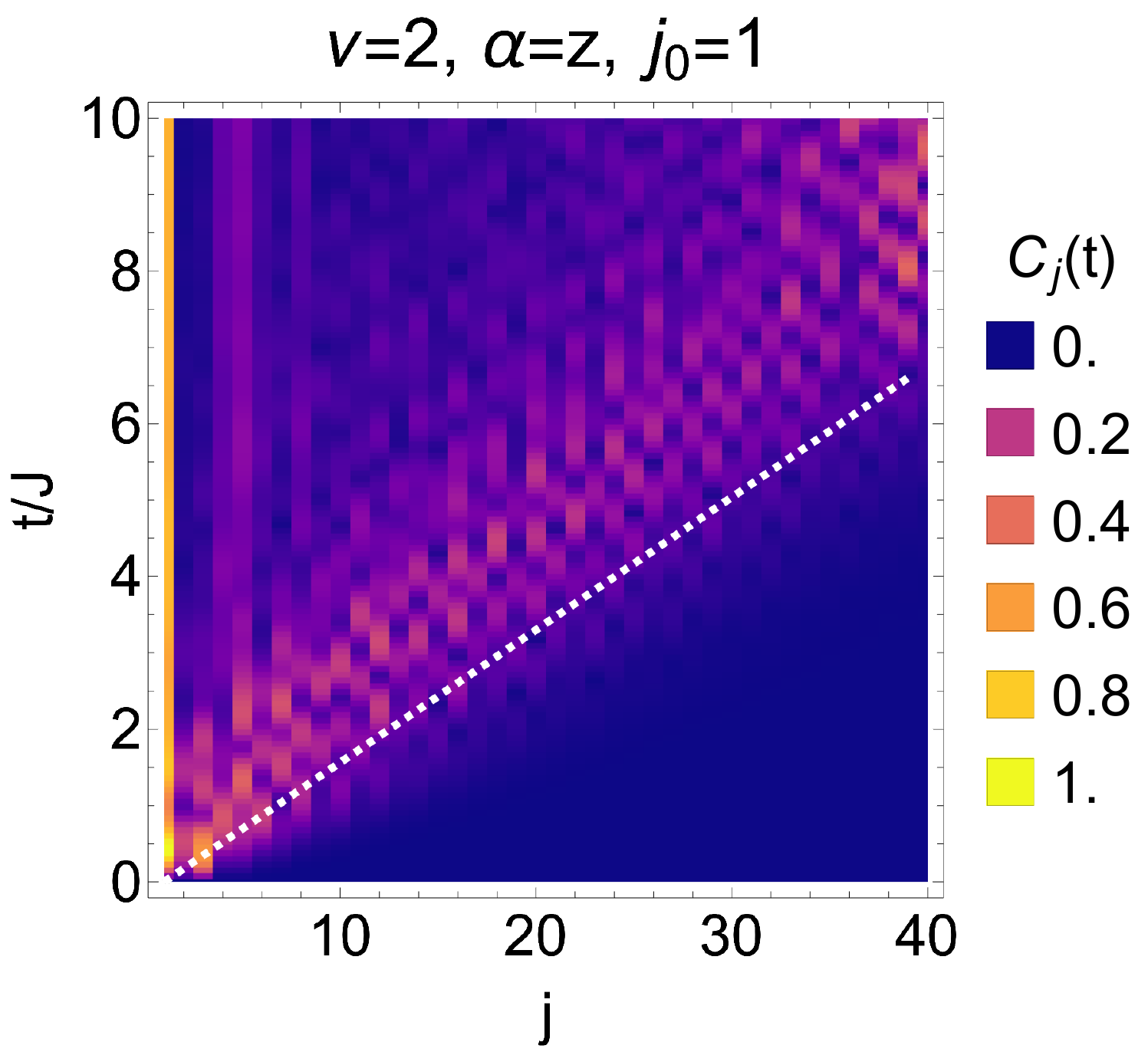}
	\includegraphics[width=0.49\columnwidth]{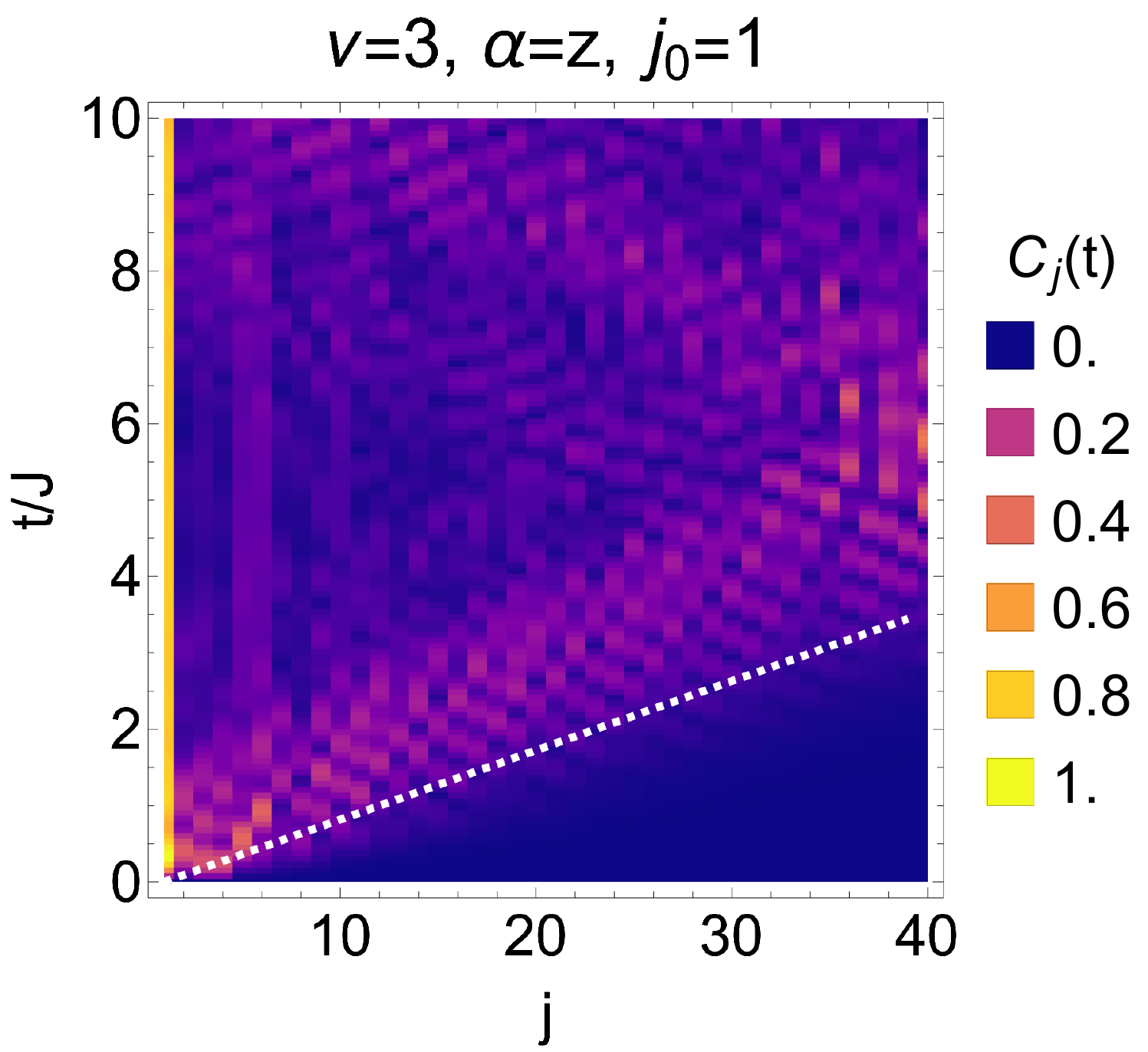}
	\caption{The OTOC $C_j(t)$ for the Kitaev model. The perturbation occurs at $j_0=20$ in the upper panels and $j_0=1$ in the lower panels and the system size is $N=40$. Results for $\hat H$ in two different topological phases are shown: $\nu=1$ and $\nu=3$. The dashed white lines are an aid to the eye for how fast the correlations spread, they show the maximum group velocity of the bulk bands. The perturbation is $\hat W_{j_0}=\exp(i W_0\hat \Psi^\dagger_{j_0}{\bm\tau}^z\hat \Psi_{j_0})$.}
	\label{figapp3}
\end{figure}

We tested a wide variety of different unitary perturbations for the Kitaev chain. Here in Fig.~\ref{figapp3} we show one example equivalent to Fig.~\ref{figlrkit} for a different perturbation, a charge density term.

%\bibliography{library}


\end{document}